\documentclass[twocolumn]{autart}    

\usepackage{amssymb,latexsym,amsfonts,amsmath}
\usepackage[utf8]{inputenc}
\usepackage{mdframed,lipsum}

\usepackage{wrapfig} 

\newenvironment{authorbio}[2][]{%
	\par\addvspace{1em}%
	\noindent
	\if\relax\detokenize{#1}\relax 
	\else
	\begin{wrapfigure}{l}{0.8in} 
		\vspace{-5pt} 
		\includegraphics[width=1in,height=1.25in,clip,keepaspectratio]{#1}
	\end{wrapfigure}%
	\fi
	\noindent\textbf{#2} 
}{\par\addvspace{2em}}

\usepackage{graphicx}

\usepackage{algorithm}
\usepackage{algorithmic}

\usepackage{mathrsfs}

\usepackage{graphicx}
\usepackage{paralist}
\usepackage{dsfont}
\usepackage{eso-pic}
\usepackage{epsfig}
\usepackage{mathtools}
\usepackage{epstopdf}

\usepackage{stfloats}
\usepackage{enumitem}

\usepackage{natbib}

\usepackage{xspace}

\usepackage{xcolor}
\definecolor{lightblue}{rgb}{0.30, 0.75, 0.93}
\definecolor{lightblue}{rgb}{0.30, 0.75, 0.93}
\definecolor{mycolor}{rgb}{0, 0.45, 0.75}
\definecolor{mycolor1}{rgb}{0.39, 0.83, 0.07}

\newcommand{\greensquare}{\tikz\fill[mycolor1!40!white] (0,0) rectangle (2mm,2mm);}
\newcommand{\redsquare}{\tikz\fill[red!30!white] (0,0) rectangle (2mm,2mm);}

\DeclareRobustCommand\sampleline[1]{%
    \tikz\draw[#1] (0,0) -- (2em,0);%
}

\usepackage{graphicx}
\usepackage{paralist}
\usepackage{dsfont}
\usepackage{caption} 
\usepackage{booktabs}
\usepackage{eso-pic}
\usepackage{epsfig}
\usepackage{mathtools}
\usepackage {url}
\usepackage{epstopdf}
\usepackage{tikz}
\usepackage{IEEEtrantools}
\usetikzlibrary{arrows,calc,decorations.markings,math,arrows.meta}
\newtheorem{theorem}{Theorem}[section]
\newtheorem{lemma}[theorem]{Lemma}
\newtheorem{proposition}[theorem]{Proposition}

\newtheorem{definition}[theorem]{Definition}
\newtheorem{example}[theorem]{Example}

\newtheorem{remark}[theorem]{Remark}

\numberwithin{equation}{section}

\usepackage[many]{tcolorbox}
\usetikzlibrary{calc}
\tcbuselibrary{skins}

\newtcolorbox{resp}[1][]{%
	enhanced jigsaw,%
	colback=gray!5!white,%
	colframe=gray!80!black,%
	size=small,%
	boxrule=1pt,%
	halign title=flush center,%
	coltitle=black,%
	breakable,%
	drop shadow=black!50!white,%
	attach boxed title to top left={xshift=1cm,yshift=-\tcboxedtitleheight/2,yshifttext=-\tcboxedtitleheight/2},%
	minipage boxed title=3cm,%
	boxed title style={%
		colback=white,%
		size=fbox,%
		boxrule=1pt,%
		boxsep=2pt,%
		underlay={%
			\coordinate (dotA) at ($(interior.west) + (-0.5pt,0)$);
			\coordinate (dotB) at ($(interior.east) + (0.5pt,0)$);
			\begin{scope}[gray!80!black]
				\fill (dotA) circle (2pt);
				\fill (dotB) circle (2pt);
			\end{scope}
		}%
	},%
	#1%
}

\newcommand{\R}{{\mathbb{R}}}

\newcommand{\EE}{\mathds{E}}
\newcommand{\PP}{\mathds{P}}

\usepackage{dsfont}

\definecolor{fluorescentpink}{rgb}{1.0, 0.08, 0.58}
\definecolor{royalblue(web)}{rgb}{0.25, 0.41, 0.88}

\makeatletter
\long\def\@maketablecaption#1#2{\@tablecaptionsize
    \global \@minipagefalse
    \hbox to \hsize{\parbox[t]{\hsize}{\centering #1 \\ #2}}}
\allowdisplaybreaks
\makeatother

\begin{document}

\begin{frontmatter}
\title{Compositional Design of Safety Controllers for Large-Scale Stochastic Hybrid Systems\thanksref{footnoteinfo}}

\thanks[footnoteinfo]{The material in this paper was
	partially presented at the American Control Conference, June 8-10, 2022, Atlanta, GA. Corresponding author:  Mahdieh Zaker (m.zaker2@newcastle.ac.uk).}

\author{Mahdieh Zaker}\ead{m.zaker2@newcastle.ac.uk},  
\author{Omid Akbarzadeh}\ead{o.akbarzadeh2@newcastle.ac.uk},
\author{Behrad Samari}\ead{b.samari2@newcastle.ac.uk},  
\author{Abolfazl Lavaei}\ead{abolfazl.lavaei@newcastle.ac.uk}  
\address{School of Computing, Newcastle University, United Kingdom}

\begin{keyword}  
Stochastic hybrid systems, scalable safety controllers, control barrier certificates, small-gain reasoning
 
\end{keyword}                        
                                         
\begin{abstract}   
In this work, we propose a compositional scheme based on {small-gain reasoning} to synthesize safety controllers for interconnected stochastic hybrid systems. In our proposed setting, we first offer an augmented scheme that characterizes each stochastic hybrid subsystem, endowed with both continuous evolution and instantaneous jumps, within a unified framework including both scenarios, implying that its state trajectories coincide with those of the original hybrid subsystem. We then introduce the concept of augmented {control sub-barrier certificates} (A-CSBCs) for each subsystem, thereby enabling the construction of an augmented {control barrier certificate} (A-CBC) for an interconnected network {(from A-CSBCs of its subsystems)} along with its safety controller under small-gain compositional conditions. We eventually leverage the constructed A-CBC {to derive a guaranteed lower bound on the safety probability of the interconnected network. While in a monolithic scheme the computational complexity of synthesizing a control barrier certificate via sum-of-squares (SOS) optimization scales polynomially with the overall network size, the proposed compositional framework reduces this dependence to the subsystem size.} We illustrate the efficacy of the proposed approach on an interconnected network comprising $1000$ stochastic hybrid subsystems with nonlinear dynamics under {two distinct interconnection topologies}.
\end{abstract}

\end{frontmatter}

\section{Introduction}
Stochastic hybrid systems~\citep{blom2006stochastic,cassandras2018stochastic} are complex heterogeneous models spanning a wide range of safety-critical applications, in which discrete dynamics describe {computational and communication components}, while continuous {dynamics} capture physical {processes}. Providing a formal verification and controller synthesis framework for such complex heterogeneous systems naturally poses significant challenges, primarily due to \emph{(i)} the stochastic nature of the dynamics, \emph{(ii)} the tight interaction between continuous and discrete components (in both space and time), \emph{(iii)} large dimensionality of the state and input sets, and \emph{(iv)} the need to manage complex logical requirements~\citep{baier2008principles}. 

In order to address the above-mentioned difficulties, one potential solution proposed in the relevant literature is to approximate the original models by simpler ones with either lower dimensionality (\emph{a.k.a.} infinite abstractions) or discrete-state sets (\emph{a.k.a.} finite abstractions)~\citep{julius2009approximations,APLS08}. On the downside, the existing abstraction-based techniques rely on discretizing state/input sets and are not applicable to real-world high-dimensional systems due to the state-explosion problem. To enhance the scalability of abstraction-based techniques, compositional approaches have been proposed in the relevant literature to construct abstractions of large-scale systems upon those of smaller subsystems~\citep{lavaei2022dissipativity,hahn2013compositional,nejati2020compositional,lavaei2022automated,lavaei2024abstraction}.

To mitigate the computational complexity arising from abstraction-based approaches, control barrier certificates (CBCs) have garnered significant attention in recent years for the formal analysis and synthesis of complex dynamical systems, offering a promising \emph{discretization-free} approach. Specifically, a barrier certificate is a Lyapunov-like function whose level set over the set of initial states separates the unsafe region from all system trajectories originating from that set. Consequently, the existence of such a function provides a (probabilistic) safety certificate for the system. Methodologies based on the notion of CBCs, initially proposed by~\citet{prajna2004safety,prajna2007framework}, have been widely employed for the formal verification and synthesis of non-stochastic~\citep{borrmann2015control,ames2019control} and stochastic dynamical systems~\citep{zhang2010safety,ahmadi2019safe,santoyo2019verification,clark2019control,nejati2022dissipativity,lavaei2024scalable}.

{Despite the extensive literature on safety controller synthesis using the notion of CBCs, to the best of our knowledge, there are no results on the compositional design of CBC-based safety controllers for large-scale stochastic hybrid systems.}
Accordingly, our primary contribution is to propose a {compositional scheme} for safety controller synthesis of interconnected stochastic hybrid systems with both continuous evolution, governed by stochastic differential equations with Brownian motions and Poisson processes, and instantaneous jumps, modeled by discrete-time stochastic reset equations with additive noise. In our proposed setting, we first present an augmented framework that recasts each stochastic hybrid subsystem with continuous evolution and instantaneous jumps via an integrated framework spanning both scenarios. We then introduce the notion of augmented \emph{control sub-barrier certificates}, computed for each subsystem, using which one can compositionally construct augmented \emph{control barrier certificates} for interconnected networks together with their safety controllers under some small-gain compositional conditions. Subsequently, we make use of the constructed augmented control barrier certificate to {derive a certified lower bound on the safety probability} of the interconnected network.

While a recent work by~\citet{lavaei2024scalable} addresses networks of discrete-time \emph{switched} systems with jumps acting solely through the transition map, our work considers a broader class of stochastic hybrid systems. These include both continuous- and discrete-time (reset) dynamics, with switching in both flow and jump components, modeled via stochastic differential and reset equations driven by Brownian motion, Poisson processes, and additive noise. This richer structure requires safety analysis for both evolution modes and necessitates additional design elements (cf. Theorem~\ref{Thm:Main}) to ensure soundness across the hybrid dynamics. Moreover, unlike the setting considered by~\citet{lavaei2024scalable}, which lacks control inputs and focuses on the design of switching signals under dwell-time constraints, our model explicitly incorporates control inputs into both the flow and jump dynamics. Accordingly, the controller synthesis problem becomes more challenging in our setting, as it involves two distinct types of {adversarial inputs}: \emph{(i)} disturbance inputs, which capture the influence of subsystems on one another, and \emph{(ii)} switching between flow and jump modes in a hybrid setting with both continuous- and discrete-time dynamics.

Compared with the related work by~\citet{nejati2022compositional}, our paper treats a more general and technically demanding class of stochastic hybrid systems. In particular, while~\citet{nejati2022compositional} model hybrid behavior through random Markovian switching in systems driven by Brownian motion and a Poisson process, our hybrid dynamics arise from state jumps occurring at constrained inter-jump times determined by jump parameters and a sampling period, which explicitly couples continuous evolution with discrete updates and additionally allows additive noise in the discrete dynamics (cf. Definition~\ref{Def:SHS}). This structural difference necessitates an integrated modeling and certification framework that can simultaneously account for both flow and jump behavior (cf. Definition~\ref{Def:A-SHS}) when designing safety controllers and composing multiple subsystems. Additionally, in our setting, safety certificates should enforce separate requirements on both continuous and discrete parts, and then reconcile them via a weighted construction that accommodates three different scenarios
(see Definition~\ref{cbc} and Theorem~\ref{Thm:Main}). In contrast, \citet[Definition~3.1]{nejati2022compositional} use a transition-rate-weighted aggregation tailored to Markovian switching with a strictly positive decay parameter. We note that while our work concerns safety controller synthesis, \citet{nejati2022compositional} focus on satisfying more challenging logical specifications expressed as deterministic finite automata.

While the initial results were presented in~\citep{lavaei2022safety}, the proposed framework is limited to monolithic systems and, due to the computational complexity of synthesizing control barrier certificates, remains practically restricted to low-dimensional models (\emph{e.g.,} around six dimensions). In contrast, we introduce an innovative compositional approach that scales to large stochastic hybrid systems (cf. the $1000$-dimensional case study). In fact, while the computational complexity of designing a CBC in~\citep{lavaei2022safety} grows \emph{polynomially} depending on the total system size when using SOS optimizations, our compositional approach reduces this dependence to the subsystem-level size (cf. Section~\ref{subsec:complexity}). Moreover,  a key challenge in analyzing large-scale interconnected stochastic hybrid systems lies not only in the high dimensionality inherent to such systems, but also in the coupling between subsystems. This coupling, modeled as {a disturbance (or adversarial input)} in each subsystem’s dynamics (cf.~\eqref{Eq:1}), poses a central difficulty specifically due to the \emph{hybrid nature of the system}, necessitating a more careful treatment of these interconnections (cf. Remark~\ref{rmk:novelty}), which is absent in~\citep{lavaei2022safety}.

\section{Stochastic Hybrid Systems}\label{Sec:SHSs}

\subsection{Notation}

Sets of real, positive and non-negative real numbers are represented by $\mathbb{R}$, $\mathbb{R}^+$, and $\mathbb{R}_{0}^+$, respectively. Sets of non-negative and positive integers are, respectively, denoted by $\mathbb{N} \coloneq \{0,1,2,...\}$ and $\mathbb{N}^+=\{1,2,...\}$. Given $N$ vectors $x_i \in \mathbb{R}^{n_i}$, $x=[x_1;\ldots;x_N]$ represents the corresponding vector of dimension $\sum_{i=1}^N n_i$. Given a vector $x\in \mathbb{R}^{n}$, we define its Hadamard power as $x^{\circ m} \coloneq [x_1^m;\ldots;x_n^m]$ for any $m \in \mathbb{N}$. {A (block) diagonal matrix with $N$ diagonal matrix entries $(A_1,\ldots,A_N)$ and $N$ scalar entries $(a_1,\ldots,a_N)$ is denoted by $\mathsf{blkdiag}(A_1,\ldots,A_N)$ and $\mathsf{diag}(a_1,\ldots,a_N)$, respectively.} 
The notation $\{a_{ij}\}$ denotes a matrix whose entry in the $i$-th row and $j$-th column is given by $a_{ij}$.
We denote the Euclidean norm of a vector $x\in\mathbb{R}^{n}$ by $\Vert x\Vert$. Given sets $X_i$, $i\in\{1,\ldots,N\}$,
their Cartesian product is denoted by $\prod_{i=1}^{N}X_i$, while their union is indicated by $\cup_{i = 1}^NX_i$. 
{The empty set is represented by $\emptyset$.} We denote the identity matrix in $\mathbb R^{n\times{n}}$ by $\mathds{I}_n$.
{Given a measurable function $f:\mathbb T\rightarrow\mathbb{R}^n$, where $\mathbb T$ is either $\mathbb{N}$ or $\mathbb{R}_0^+$, the (essential) supremum of $f$ is denoted by $\|f\|_{\infty} \coloneq \operatorname*{(ess)\,sup}_{t\in\mathbb{T}} \|f(t)\|$.}

\subsection{Preliminaries}
Given a probability space $(\Omega,\mathcal F_{\Omega},\mathds{P}_{\Omega})$, {where} $\Omega$ is the sample space,
$\mathcal F_{\Omega}$ is a sigma-algebra on $\Omega$ {including subsets of $\Omega$ as events}, and $\mathds{P}_{\Omega}$ is a probability measure {assigning probabilities to events}, we assume that $(\Omega,\mathcal F_{\Omega},\mathds{P}_{\Omega})$ is endowed with a filtration $\mathds{F} = (\mathcal F_s)_{s\geq 0}$ fulfilling conditions of completeness and right continuity. We consider $(\mathbb W_s)_{s \ge 0}$ as a ${\textsf b}$-dimensional $\mathds{F}$-Brownian motion and $(\mathbb P_s)_{s \ge 0}$ as an ${\textsf r}$-dimensional $\mathds{F}$-Poisson process. The Poisson process and Brownian motion are considered to be mutually independent. The Poisson process $\mathbb P_s = [\mathbb P_s^1; \ldots; \mathbb P_s^{\textsf r}]$ models ${\textsf r}$ events with independent occurrences. {We assume for each $\bar{z} \in \{1, \ldots, \textsf r\}$, $\mathbb P_s^{\bar{z}}$ has a rate $\lambda_{\bar{z}}$.}

\subsection{Stochastic Hybrid Systems}
In this work, our focus lies on  (a class of) stochastic hybrid systems, treating them as {individual subsystems}. We begin by formally introducing
it, which generalizes the impulsive system concept in~\citep[Definition~1]{swikir2020symbolic} by including different sources of noise and disturbances.

\begin{definition}\label{Def:SHS}
	A stochastic hybrid system (SHS) $\Phi_i$ is characterized by $\Phi_i\!=\!(X_i, U_i,\mathcal{U}_i,W_i,\mathcal{W}_i,\sigma_i,\rho_i,f_{1_i},$ $\varsigma_i,f_{2_i})$,
	where: 
	\begin{itemize}
		\item $X_i\subseteq\mathbb{R}^{n_i}$ is the state set of $\Phi_i$;
		\item $U_i\subseteq\mathbb{R}^{m_i}$ is the input set of $\Phi_i$;
		{\item $\mathcal U_i$ is a subset of all bounded $\mathds F$-progressively measurable {processes taking values in $U_i$};}
		\item $W_i\subseteq\mathbb{R}^{p_i}$ is the disturbance set of $\Phi_i$ (which is utilized later for the sake of interconnection);
		\item {$\mathcal W_i$ is a subset of all bounded $\mathds F$-progressively measurable c\`adl\`ag {processes taking values in $W_i$};}
		\item  $\sigma_i: {X_i} \rightarrow \mathbb R^{n_i\times \textsf{b}_i}$ is the diffusion term which is globally Lipschitz continuous;
		\item $\rho_i: {X_i} \rightarrow \mathbb R^{n_i\times \textsf{r}_i}$ is the reset term which is globally Lipschitz continuous;
		\item $f_{1_i}: X_i\times  U_i \times W_i\rightarrow {\R^{n_i}}$ is the drift term which is globally Lipschitz continuous;
		\item $\varsigma_i$ is a sequence of independent and identically distributed
		(i.i.d.) random variables {(independent of the continuous-time noise sources)} from the sample space $\Omega_i$ to the measurable space $(\mathcal{V}_{\varsigma_i},\mathcal F_{\varsigma_i})$:	 
		\begin{equation*}
			\varsigma_i\coloneq\big\{\varsigma_i(t) : (\Omega_i,\mathcal F_{\Omega_i})\rightarrow (\mathcal{V}_{\varsigma_i},\mathcal F_{\varsigma_i}),\; t\in\Delta_i\big\},
		\end{equation*}
		{where {$\Delta_i=\{t_k\}_{k\in\mathbb N}$} with $t_{k+1}-t_{k}\in\{\varepsilon_{1_i}\tau,\ldots,\varepsilon_{2_i}\tau\}$ for jump parameters $\tau\in\mathbb R^+$ and $\varepsilon_{1_i},\varepsilon_{2_i}\in \mathbb N^+$, $\varepsilon_{1_i}\le \varepsilon_{2_i}$;}
		\item $f_{2_i}: X_i\times U_i \times W_i\times \mathcal V_{\varsigma_i}\rightarrow X_i$ is the transition map which is globally Lipschitz continuous.
	\end{itemize}
	The SHS $\Phi_i$ {satisfies}
	\begin{align}\label{Eq:1}
		\Phi_i\!:\left\{\hspace{-1.5mm}
		\begin{array}{rl}
			\mathsf{d}x_i(t)\!=\!\!& f_{1_i}(x_i(t),\nu_i(t),w_i(t))\mathsf{d}t\!+\!\sigma_i(x_i(t))\mathsf{d}\mathbb W_t\\
			&+\rho_i(x_i(t))\mathsf{d}\mathbb P_t,~\quad\quad\quad\quad\quad\quad~~~\!\!\!\! {t\in\mathbb{R}_{0}^+\backslash \Delta_i},\\
			x_i(t)\!=\!\!& f_{2_i}(x_i(t^-),\nu_i(t),w_i(t^-),\varsigma_i(t)),\quad \!\!\! {t \in \Delta_i},
		\end{array}
		\right.
	\end{align}
	{$\mathds P$-almost surely ($\mathds P$-a.s.) for any $\nu_i \in \mathcal{U}_i$ and any $w_i \in \mathcal{W}_i$, where the stochastic process $x_i : \Omega_i\times\mathbb R_{0}^+\rightarrow X_i$ is referred to as the solution process of $\Phi_i$ (the processes $x_i$ and $w_i$ are assumed to be right-continuous and to have left limits for all $t\in\mathbb R_{0}^+$). We denote by $x_{x_{0_i}\nu_i w_i}(t)$ the value of the solution process at time $t \in \mathbb{R}_{0}^+$ under input and disturbance trajectories $\nu_i$ and $w_i$\footnote{With a slight abuse of notation, the symbols $x_i$, $\nu_i$, and $w_i$ are used to denote both the {processes} and the values of the state, input, and disturbance; the intended meaning will be clear from the context.} from an initial condition $x_{0_i}\in X_i$, $\mathds P$-a.s., where $x_{0_i}$ is a random variable that is $\mathcal F_0$-measurable. Moreover, $x_i(t^-)\coloneq \lim_{s\nearrow t} x_i(s)$ and $w_i(t^-)\coloneq \lim_{s\nearrow t} w_i(s)$ denote the values of $x_i$ and $w_i$ as $s$ approaches $t$ from the left.}
\end{definition}
\begin{remark}
	In the SHS defined in Definition~\ref{Def:SHS}, jumps at time instants {$t \in \Delta_i$} introduce discontinuities in the state trajectories $x_i(t)$, which evolve discretely at such times. To ensure the well-posedness of the model, it is necessary to specify the left-hand limit $x_i(t^-)$ at each jump instant. The same requirement applies to the {disturbance} $w_i(t)$, representing the (possibly discontinuous) state trajectories of neighboring subsystems (cf. \eqref{Eq:41}). In contrast, the control input $\nu_i(t)$ is user-designed and can be applied as a continuous-time signal without needing to exhibit jumps~\citep{swikir2020symbolic}. The additive noise $\varsigma_i(t)$ is also defined only at jump instants $t \in \Delta_i$ and hence does not require evaluation at $t^-$.
\end{remark}

{\begin{remark}
	The global Lipschitz assumptions on the drift, diffusion, reset functions, and the transition map of the SHS $\Phi_i$ are imposed to guarantee well-posedness of the subsystem dynamics. It is worth noting that, since our CBC analysis is confined to compact domains, local Lipschitz continuity on an open neighborhood of these domains is typically sufficient.
\end{remark}}

To further clarify the modeling capabilities of such systems and the roles of the Brownian motion and Poisson process, we present the following illustrative example.
\begin{example}
		Consider a network of $N$ rooms, each equipped with a local heater. The dynamics in~\eqref{Eq:1} can model each room, where $x_i$ denotes its temperature, controlled by the heater control input $\nu_i$, and influenced by $w_i$, the inter-room heat transfer (e.g., through shared walls). The Brownian motion $\mathbb{W}_t$ models continuous environmental uncertainties (e.g., brief door/window openings), while the Poisson process $\mathbb{P}_t$ captures discrete events such as people entering, causing sudden temperature changes. The jump times {$\Delta_i = \{t_k\}_{k \in \mathbb{N}}$} correspond to sudden thermostat reconfigurations (e.g., switching between ECO and COMFORT modes), occurring at intervals constrained by $\varepsilon_{1_i}\tau \leq t_{k+1} - t_k \leq \varepsilon_{2_i}\tau$.
\end{example}

Given that control inputs in practice are often fed by a digital controller, we are interested here in sampled-data hybrid systems in which the control inputs are considered to be piecewise constant as
\begin{align*}
	&\mathcal{U}_{\tau_i}\!\!=\!\big\{\nu_i\!:\mathbb R_{0}^+\!\rightarrow\! U_i\,\big|\, \nu_i(t)\!=\!\nu_i({(k-1)\tau}), \\
	& \hphantom{\mathcal{U}_{\tau_i}\!\!=\!\big\{} t\!\in\! [{(k-1)\tau, k \tau }), \, k \in \mathbb N^+\!\big\}.
\end{align*}

\subsection{Interconnected SHS}

Since the ultimate goal of this work is to synthesize a safety controller for an interconnected SHS, we next provide a formal definition of the interconnected SHS without the $w_i$, which can be constructed as a composition of multiple SHSs with {disturbance inputs}. 

\begin{definition}\label{Def:3}
	Consider an interconnected SHS comprising $\Phi_i =(X_i, U_i,$ $\mathcal{U}_i, W_i,\mathcal{W}_i,\sigma_i,\rho_i, f_{1_i},\varsigma_i,f_{2_i})$, $i\in\{1,\ldots,N\}$, with their disturbances partitioned as	
		\begin{align}\label{Eq:31}
			w_i&=[{w_{i1};\ldots;w_{i(i-1)};w_{i(i+1)};\ldots;w_{iN}}],
		\end{align}
		satisfying the interconnection constraint
		\begin{align}\label{Eq:41}
			\forall i,j\in \{1,\dots,N\},j\neq i\!: ~~ w_{ij}=x_{j}, ~~ X_{j}\subseteq W_{ij},
		\end{align}
		where $W_i\coloneq\prod_{j\neq i}W_{ij}$.
		Then, the interconnection of $\Phi_i$, $i \in \{1, \ldots, N\}$, forms the interconnected SHS $\Phi=(X, U,\sigma,\rho,f_1,\varsigma,f_2)$, denoted by
		$\mathcal{I}(\Phi_1,\ldots,\Phi_N)$, where $X = \prod_{i = 1}^N X_i$, $U = \prod_{i = 1}^N U_i$,
		$\sigma\coloneq\mathsf{blkdiag}(\sigma_1(x_1),$ $\ldots,\sigma_N(x_N))$, $\rho\coloneq\mathsf{blkdiag}(\rho_1(x_1),\ldots,\rho_N(x_N))$, $\varsigma\coloneq[\varsigma_1;\dots;\varsigma_N],$ $f_1\coloneq[f_{1_1};\ldots;f_{1_N}],$ and $f_2\coloneq[\bar f_{2_1};\ldots;\bar f_{2_N}],$ with
		\begin{align*}
			&\bar f_{2_i}\big((x_i(t^-),\nu_i(t),w_i(t^-),\varsigma_i(t)\big)\\
			&=\begin{cases}
				x_i(t^-), & t\notin \Delta_i,\\
				f_{2_i}\big((x_i(t^-),\nu_i(t),w_i(t^-),\varsigma_i(t)\big), & t\in\Delta_i.
			\end{cases}
		\end{align*}
		Such an interconnected SHS can be described by
		\begin{align}\label{Eq:51}
			\Phi\!:\left\{\hspace{-1.5mm}
			\begin{array}{rl}
				\mathsf{d}x(t)=\!\!& f_1(x(t),\nu(t))\mathsf{d}t+\sigma(x(t))\mathsf{d}\mathbb W_t+\rho(x(t))\mathsf{d}\mathbb P_t,\\
				&~\quad\quad\quad\quad\quad\quad\quad\quad\quad\quad\quad\quad\quad  ~t\in\mathbb{R}_{0}^+\backslash \Delta,\\
				x(t)=\!\!& f_2(x(t^-),\nu(t),\varsigma(t)),\,\quad\quad\quad\quad\quad ~\!\!t\in \Delta,
			\end{array}
			\right.
		\end{align}
		where $x = [x_1; \ldots; x_N]$, $\nu = [\nu_1; \ldots; \nu_N]$, $\Delta =\cup_{i = 1}^N\Delta_i$, while $\mathbb W_t$ and $\mathbb P_t$ are obtained by stacking the corresponding processes of the subsystems.
	The value of the solution process of $\Phi$ at time $t \in \mathbb R_{0}^+$ under an input trajectory $\nu$ starting from $x_{0}$ is denoted by $x_{x_{0}\nu}(t)$.
\end{definition}

We now formally define the safety specification of interest for the interconnected SHS.
\begin{definition}\label{Def:safety}
	Given a safety specification $\Lambda = (X_{0}, X_{u}, \mathcal{T})$, with $X_{0}, X_{u} \subset X$ representing, respectively, the initial and unsafe sets of the interconnected SHS $\Phi$, where $X_{0} \cap X_{u} = \emptyset$, the network $\Phi$ is said to be safe within the time horizon $\mathcal{T} \in \mathbb{N}$, denoted by $\Phi \vDash \Lambda$, if all trajectories starting from $X_{0}$ avoid $X_{u}$ during $\mathcal{T}$. Since the trajectories of $\Phi$ are stochastic, we are interested in computing the probability $\mathds{P}\{\Phi \vDash \Lambda\} \geq 1 - \epsilon$, for some $\epsilon \in (0,1]$.
\end{definition}

\subsection{Augmented SHS}
As can be observed, the SHS in~\eqref{Eq:1} encompasses both continuous and discrete evolutions, in which continuous behaviors are governed by stochastic differential equations with Brownian motions and Poisson processes, and instantaneous jumps are modeled by discrete-time stochastic reset equations with additive noise. To enable a \emph{unified treatment} of these two modes of evolution, and to handle the logic governing jump timing, we introduce an augmented representation of SHS, adopted from~\citet[Definition~5]{swikir2020symbolic} but adapted to the stochastic setting, as formalized in the following definition.

\begin{definition}\label{Def:A-SHS}
	Consider an SHS $\Phi_i=(X_i, U_i,\mathcal{U}_i,W_i,\\\mathcal{W}_i,\sigma_i,\rho_i, f_{1_i},\varsigma_i,f_{2_i})$ with jump parameters ($\tau$, $\varepsilon_{1_i}$, $\varepsilon_{2_i}$). An augmented SHS (A-SHS) $\mathcal{A}_i(\Phi_i)$ corresponding to $\Phi_i$ is characterized by {$\mathcal{A}_i(\Phi_i)\!=\!(\mathbb X_i,\mathbb U_i,\mathscr U_i,\mathds W_i,\mathscr W_i,\sigma_i,\rho_i,\varsigma_i,$ $\mathbb F_i,\mathbb Y_i,\mathbb H_i)$}, where:
	\begin{itemize}
		\item  $\mathbb X_i=X_i \times \{0,\ldots,\varepsilon_{2_i}\}$ is the state space of $\mathcal{A}_i(\Phi_i)$, where $(x_i,\vartheta_i) \in \mathbb{X}_i$ indicates that the current state is $x_i$, and the time elapsed since the latest jump is $\vartheta_i$;
		\item $\mathbb U_i=U_i$ is the input space of $\mathcal{A}_i(\Phi_i)$;
		{\item $\mathscr U_i = \mathcal{U}_{\tau_i}$;}
		\item {$\mathds W_i=W_i$} is the disturbance space of $\mathcal{A}_i(\Phi_i)$;
		\item $\mathscr W_i$ is the set of disturbance process segments over intervals of length $\tau$, represented on $[0,\tau)$, associated with $\mathcal W_i$;
		\item $\sigma_i$ is the diffusion term;
		\item $\rho_i$ is the reset term;
		\item $\varsigma_i$ is a sequence of i.i.d. random variables;
		\item  {$(x_i',\vartheta_i')\in \mathbb F_i((x_i,\vartheta_i),\nu_i,w_i,\sigma_i,\rho_i,\varsigma_i)$} if and only if one of the following two cases holds:
		\begin{enumerate}[label=(\roman*)]
			\item Flow case: 
   $0\leq \vartheta_i\leq \varepsilon_{2_i}-1$,\\ $x_i'=  x_{x_i\nu_iw_i}(\tau^-)$, and $\vartheta_i'=\vartheta_i+1$;
			\item Jump case: $\varepsilon_{1_i}\leq \vartheta_i\leq \varepsilon_{2_i}$,\\ $x_i'= {f_{2_i}(x_i,\nu_i(0),w_i(0),\varsigma_i(0))}$, and $\vartheta_i'=0$. 
		\end{enumerate}
		The randomness induced by the Brownian motion and Poisson process is captured implicitly through the solution process. In addition, $w_i$ denotes a segment of the disturbance process over an interval of length $\tau$, represented on $[0,\tau)$, which is assumed to remain bounded on this interval.
		\item {$\mathbb Y_i=X_i$} is the output space of $\mathcal{A}_i(\Phi_i)$;
		\item $\mathbb H_i:\mathbb X_i\rightarrow \mathbb Y_i$ is the output map defined as $\mathbb H_i(x_i,\vartheta_i)=x_i$.
	\end{itemize}
\end{definition}

\begin{remark}
	Note that {$(x_i',\vartheta_i')\in\mathbb F_i((x_i,\vartheta_i),\nu_i,w_i,\sigma_i,\\\rho_i,\varsigma_i)$} in Definition~\ref{Def:A-SHS} characterizes the evolution of the A-SHS $\mathcal{A}_i(\Phi_i)$, in which $\mathbb {F}_i$ is acquired based on one of the following two cases:\vspace{0.15cm}\\
	$\bullet$ if the counter $\vartheta_i$ is between $0$ and $\varepsilon_{2_i}-1$, then the state may evolve based on the flow case as $x_i' =  x_{x_i\nu_iw_i}(\tau^-)$, in which $x_{x_i\nu_iw_i}(\tau^-)$ is the solution of the stochastic differential equation  $\mathsf{d}x_i(t)=f_{1_i}(x_i(t),\nu_i(t),w_i(t))\mathsf{d}t\!+\!\sigma_i(x_i(t))\mathsf{d}\mathbb W_t+\rho_i(x_i(t))\mathsf{d}\mathbb P_t,$~\citep{oksendal2005applied}. In this case, the next counter is increased as $\vartheta_i'=\vartheta_i+1$;\vspace{0.15cm}\\
	$\bullet$ if the counter $\vartheta_i$ is between $\varepsilon_{1_i}$ and $\varepsilon_{2_i}$, then the state may evolve based on the jump case as $x_i'=f_{2_i}(x_i,\nu_i(0),w_i(0),\varsigma_i(0))$ and the counter is reset to zero as $\vartheta_i'=0$.\\
	{In other words, at each transition step of $\mathcal{A}_i(\Phi_i)$, depending on the counter value, the subsystem may either execute a flow transition over one sampling interval or execute a jump transition immediately from the current state. This is the main reason of taking the values of $\nu_i(0)$, $w_i(0)$, and $\varsigma_i(0)$, denoting their current values for the current run. The counter records how many sampling intervals have elapsed since the last jump and determines when jump transitions are enabled or forced.}
\end{remark}
We further clarify the evolution of the SHS and the A-SHS in Definitions~\ref{Def:SHS} and~\ref{Def:A-SHS}, respectively, and how they are related to each other through the following illustrative example.
\begin{example}\label{ex:SHS}
	Consider the case where $\varepsilon_{1_i}=1$ and $\varepsilon_{2_i}=3$. Under Definition~\ref{Def:SHS}, the system evolves continuously according to the stochastic differential equation until a jump occurs. If the first jump happens at time $t_k$, then the next jump at $t_{k+1}$ must satisfy $t_{k+1} - t_k \in \{\tau, 2\tau, 3\tau\},$ meaning that successive jumps occur no sooner than $\tau$ and no later than $3\tau$ after the preceding jump. If no jump takes place at $\tau$ or $2\tau$ following $t_k$, a jump is guaranteed at $3\tau$. This pattern repeats for every pair of consecutive jumps. In Definition~\ref{Def:A-SHS}, the {auxiliary} state variable $\vartheta_i$ serves as a {counter}, storing the elapsed time between consecutive jumps, and increments by one at each time step of length $\tau$. Hence, when $\varepsilon_{1_i}=1$ and $\varepsilon_{2_i}=3$, while $\vartheta_i$ (an integer value) remains within the inclusive range $[\varepsilon_{1_i},\varepsilon_{2_i}]$, a jump \emph{may occur}, specifically at $\vartheta_i=1$ (after $\tau$) or at $\vartheta_i=2$ (after $2\tau$). If no jump happens by the time $\vartheta_i$ reaches $\varepsilon_{2_i}=3$, a jump is then enforced at $3\tau$. Immediately following any jump, $\vartheta_i$ {resets to zero}, beginning a new count toward the next permissible jump interval.
	More precisely, the system begins in the flow case with $\vartheta_i=0$, after which the counter increments to one ($\vartheta_i=1$). Then, since $\varepsilon_{1_i}\leq\vartheta_i\leq\varepsilon_{2_i}$, a jump \emph{might} happen, and if so, the system evolves following the jump case {(the state value changes)}. Otherwise, $\vartheta_i$ increases to two ($\vartheta_i=2$). If a jump does not happen again, $\vartheta_i$ becomes three ($\vartheta_i=3$). Since $0\leq\vartheta_i\leq\varepsilon_{2_i}-1$ does not hold anymore, the system necessarily evolves according to the jump case, and $\vartheta_i$ resets to zero to count the time elapsed until the next jump. Thereby, Definitions~\ref{Def:SHS} and~\ref{Def:A-SHS} represent the same behavior.
\end{example}
While Definitions~\ref{Def:SHS} and \ref{Def:A-SHS} are clearly connected through Example~\ref{ex:SHS}, the following proposition  establishes that the state trajectories of an A-SHS and its original SHS coincide at $\tau$-length instants. Consequently, we leverage the A-SHS as a unified framework encompassing both continuous evolution and instantaneous jumps, providing a more manageable model.

\begin{proposition}\label{Proposition}
	For any fixed initial condition $\bar x_{0_i} = (x_{0_i}, \vartheta_{0_i})$ (with $\vartheta_{0_i} = 0$ being the initial counter of the A-SHS), the same input and disturbance {trajectories} $\nu_i(\cdot)$ and $w_i (\cdot)$, and the same realizations of the stochastic processes $\mathbb W_t,\mathbb P_t,$ and the random sequence $\varsigma_i(\cdot),$ the {solution process} of $\Phi_i$, as presented in Definition~\ref{Def:SHS},
	coincides with a state trajectory of $\mathcal{A}_i(\Phi_i)$ in Definition~\ref{Def:A-SHS} at time instants $k\tau, k \in \mathbb{N}$, and vice versa.
\end{proposition}

\section{Augmented Control (Sub-)Barrier Certificates}\label{sec:SPSF}

In this section, we introduce the notions of augmented control sub-barrier and barrier certificates for augmented SHSs with and without {disturbance inputs}, respectively. We then utilize these notions to quantify a guaranteed probabilistic bound on the safety of the interconnected SHS.

The following definition formalizes the notion of an augmented control sub-barrier certificate, which acts as a local certificate for each subsystem, borrowed from~\citet[Definition~3.3]{lavaei2022formal}, but adapted to the SHS setting.

\begin{definition}\label{Def_1a1} 
	Given an A-SHS {$\mathcal{A}_i(\Phi_i)=(\mathbb X_i,\mathbb U_i,\mathscr U_i,$ $\mathds W_i,\mathscr W_i,\sigma_i,\rho_i,\varsigma_i,\mathbb F_i,\mathbb Y_i,\mathbb H_i)$}, consider $\mathbb X_{0_i}=X_{0_i}\times \{0\}$, and $\mathbb{X}_{u_i} = X_{u_i} \times \{0,\ldots,\varepsilon_{2_i}\}$ as initial and unsafe regions of $\mathcal{A}_i(\Phi_i)$, respectively, with $X_{0_i}, X_{u_i} \subset  X_i$ being initial and unsafe regions of $\Phi_i$, respectively.
	A function $\mathcal B_i:\mathbb{X}_i\to\R_{0}^+$ is said to be an augmented control sub-barrier certificate (A-CSBC) for $\mathcal{A}_i(\Phi_i)$ if there exist constants $0<\gamma_i<1$, $\kappa_i\in\R^+,  \mu_i,\beta_i,\varphi_{i},\eta_i\in\R_{0}^+$, such that
	\begin{subequations}\label{Eq_2a}
		\begin{align}\label{Eq_2a22}
			&\mathcal B_i(x_i,\vartheta_i) \geq \kappa_i\Vert x_i\Vert^2,\quad\quad\quad\quad~\! \forall(x_i,\vartheta_i)\in \mathbb{X}_{i},\\\label{Eq_2a11}
			&\mathcal B_i(x_i,\vartheta_i) \leq \mu_i,\quad\quad\quad\quad\quad\quad\quad\!\!\!\! \forall(x_i,\vartheta_i)\in \mathbb{X}_{0_i},\\\label{Eq_2a21}
			&\mathcal B_i(x_i,\vartheta_i) \geq \beta_i, \quad\quad\quad\quad\quad\quad\quad\!\!\! \forall(x_i,\vartheta_i)\in \mathbb{X}_{u_i}, 
		\end{align}  
		and $\forall(x_i,\vartheta_i)\in \mathbb{X}_i$, $\exists \nu_i\in \mathbb U_i$, such that $\forall w_i\in \mathds{W}_i$,
		\begin{align}\label{Eq_3a1}
			\EE \Big[\mathcal B_i(x_i',\vartheta_i')\big|x_i,\nu_i,w_i,\vartheta_i\Big]\!\leq\!\gamma_i \mathcal B_i(x_i,\vartheta_i)+\varphi_{i}\Vert w_i\Vert^2+\eta_i,
		\end{align}
	\end{subequations}
	{where $(x_i',\vartheta_i')\in \mathbb F_i((x_i,\vartheta_i),\nu_i,w_i,\sigma_i,\rho_i,\varsigma_i)$}, and $\EE$ is the expected value w.r.t. {the underlying randomness in the system} under the one-step transition of $\mathcal{A}_i(\Phi_i)$.
\end{definition}
In this definition,~\eqref{Eq_3a1} expresses how the expected value of the certificate evolves under one-step transitions. The presence of $\varphi_i$ reflects the influence of interconnections, which becomes crucial in compositional safety analysis.
\begin{remark}
The order of the quantifiers in condition~\eqref{Eq_3a1}, namely $\forall x_i \in X_i$, $\exists \nu_i \in U_i$, and $\forall w_i \in W_i$, inherently facilitates the synthesis of fully decentralized controllers for $\mathcal{A}_i(\Phi_i)$. This is because the control input $\nu_i$ is independent of the adversarial input $w_i$, which represents the state information of other subsystems. As a result, this independence offers considerable flexibility in designing controllers for large-scale networks.
\end{remark}

We now present the notion of augmented control barrier certificates for augmented SHSs without {disturbance inputs}, \emph{i.e.}, interconnected A-SHSs (cf. Definition~\ref{interconnected A-SHS}). 

\begin{definition}\label{Def_1a} 
	Given an interconnected A-SHS $\mathcal{A}(\Phi)=(\mathbb X,\mathbb U,\sigma,\rho,\varsigma, \mathbb F,\mathbb Y,\mathbb H)$ with initial and unsafe regions $\mathbb X_0,\mathbb{X}_u$, a function $\mathcal B:\mathbb{X}\to\R_{0}^+$ is said to be an augmented control barrier certificate (A-CBC) for $\mathcal{A}(\Phi)$ if there exist constants $0<\gamma<1$, $\mu,\beta,\eta\in\R_{0}^+$, with $\beta > \mu$, such that
	\begin{subequations}\label{Def_1aa}
		\begin{align}\label{Eq_2a1}
			&\mathcal B(x,\vartheta) \leq \mu,\quad\quad\quad\quad\quad\quad\quad\quad\!\!\! \forall(x,\vartheta)\in \mathbb{X}_0,\\\label{Eq_2a2}
			&\mathcal B(x,\vartheta) \geq \beta, \quad\quad\quad\quad\quad\quad\quad\quad \!\!\! \forall(x,\vartheta)\in \mathbb{X}_u, 
		\end{align}  
		and $\forall(x,\vartheta)\in \mathbb{X}$, $\exists \nu\in \mathbb U$, such that
		\begin{align}\label{Eq_3a}
			\EE \Big[\mathcal B(x',\vartheta')\,\big|\,x,\nu,\vartheta\Big]\leq \gamma \mathcal B(x,\vartheta)+\eta,
		\end{align}
        {where $(x',\vartheta')\in \mathbb F((x,\vartheta),\nu,\sigma,\rho,\varsigma)$.}
	\end{subequations}
\end{definition}

We note that the condition $\beta > \mu$ is required in Definition~\ref{Def_1a} to ensure that the initial and unsafe sets of the interconnected A-SHS do not intersect. The next theorem, borrowed from~\citep{1967stochastic} but adapted to the A-SHS setting, demonstrates the usefulness of an A-CBC by providing a guaranteed probabilistic bound on the safety of the interconnected SHS.

\begin{theorem}\label{Kushner}
	Given an interconnected A-SHS $\mathcal{A}(\Phi)=(\mathbb X,\mathbb U,\sigma,\rho,\varsigma, \mathbb F,\mathbb Y,\mathbb H)$, let $\mathcal B$ be an A-CBC for $\mathcal{A}(\Phi)$. Then, for any {$(x_0,\vartheta_0)\in\mathbb X_0$}, the probability that the state trajectories of $\mathcal{A}(\Phi)$ reach the unsafe region $\mathbb X_u$ over the finite-time horizon $k\in [0,1, \dots, \mathcal T]$ is {upper-bounded} as
	{\begin{align}\notag
		&\PP \Big\{\sup_{0\leq k\leq \mathcal T}\mathcal B(x(k\tau),\vartheta(k\tau))\geq \beta\,\, \big|\,\, x_0,\vartheta_0\Big\}\\\label{eqlemma2}
		&\leq \epsilon = \begin{cases} 
			1-(1-\frac{\mu}{\beta})(1-\frac{\eta}{\beta})^{\mathcal T}\!, \quad\quad & \text{if } \beta \geq \frac{\eta}{1-{\gamma}}, \\
			(\frac{\mu}{\beta})\gamma^{\mathcal T}+(\frac{\eta}{(1-{\gamma})\beta})(1-\gamma^{\mathcal T}), \quad\quad & \text{if } \beta< \frac{\eta}{1-{\gamma}}.  \\
		\end{cases}
	\end{align}	}
	If $\eta = 0$ in~\eqref{Eq_3a} (i.e., $\mathcal B$ being a nonnegative supermartingale), the guarantee in~\eqref{eqlemma2} can be extended to an infinite-time horizon as
	{\begin{align}\label{infinite}
		&\PP \Big\{\sup_{k\geq 0}\mathcal B(x(k\tau),\vartheta(k\tau))\geq \beta\,\, \big|\,\, x_0,\vartheta_0\Big\}\leq\epsilon= \frac{\mu}{\beta}.
	\end{align}}
\end{theorem}

Searching for CBCs for large-scale dynamical systems is, in general, computationally significantly expensive. This has been the main motivation in this work to obtain an A-CSBC for each $\mathcal{A}_i(\Phi_i)$ instead of directly searching for an A-CBC for the interconnected A-SHS, which is not possible in practice. In the next section, we propose a compositional approach based on small-gain reasoning to construct an A-CBC for the interconnected A-SHS based on A-CSBCs of its individual subsystems.

\section{Compositional Construction of A-CBC}\label{Compositional}
\begin{figure*}[!t]
	\rule{\textwidth}{0.1pt}
	\begin{align}\notag
		\EE &\Big[\mathcal B(x',\vartheta')\,\big|\,x,\nu,\vartheta\Big] \overset{\eqref{Overall B}}{=} \EE \Big[\sum_{i=1}^N\xi_i\mathcal B_i(x'_i,\vartheta'_i)\,\big|\,x,\nu,\vartheta\Big] = \sum_{i=1}^N\xi_i\EE \Big[\mathcal B_i(x'_i,\vartheta'_i)\,\big|\,x,\nu,\vartheta\Big]  = \sum_{i=1}^N\xi_i\EE \Big[\mathcal B_i(x'_i,\vartheta'_i)\,\big|\,x_i,\nu_i,w_i,\vartheta_i\Big]\\\notag
		&\!\!\!\overset{\eqref{Eq_3a1}}{\leq}\sum_{i=1}^N\xi_i\big(\gamma_i \mathcal B_i(x_i,\vartheta_i)+\varphi_{i}\Vert w_i\Vert^2+\eta_i\big)\overset{\eqref{Eq:31}}{\vphantom{\leq}=\vphantom{\leq}}\sum_{i=1}^N\xi_i\big(\gamma_i \mathcal B_i(x_i,\vartheta_i)+\varphi_{i}\sum_{\substack{j=1\\i\neq j}}^N\Vert w_{ij}\Vert^2+\eta_i\big)\\\notag
		&\overset{\eqref{Eq:41}}{\vphantom{\leq}=\vphantom{\leq}}\sum_{i=1}^N\xi_i\big(\gamma_i \mathcal B_i(x_i,\vartheta_i)+\varphi_{i}\sum_{\substack{j=1\\i\neq j}}^N\Vert x_{j}\Vert^2+\eta_i\big)\overset{\eqref{Eq_2a22}}{\leq}\sum_{i=1}^N\xi_i\big(\gamma_i \mathcal B_i(x_i,\vartheta_i)+\sum_{\substack{j=1\\i\neq j}}^N\frac{\varphi_{i}}{\kappa_j}\mathcal B_j(x_j,\vartheta_j)+\eta_i\big)\\\notag
		&=\sum_{i=1}^{N}\xi_i\mathcal B_i(x_i,\vartheta_i) \!+\! \sum_{i=1}^N\xi_i\big(-\theta_i\mathcal B_i(x_i, \vartheta_i)+\sum_{\substack{j=1\\i\neq j}}^N\psi_{ij}\mathcal B_j(x_j,\vartheta_j)+\eta_i\big)\\\tag{4.3}\label{Eq:23}
		&=\mathcal B(x,\vartheta) + \xi^\top(-\Theta+\Psi)\underbrace{\big[\mathcal B_1(x_1,\vartheta_1);\ldots;\mathcal B_N(x_N,\vartheta_N)\big]}_{\tilde {\mathcal B}(x,\vartheta)}+\sum_{i=1}^N\xi_i\eta_i\overset{\eqref{Eq:21}}{\leq} \underbrace{(1 + \Gamma)}_{\gamma} \mathcal B(x,\vartheta) + \eta.
	\end{align}
	\rule{\textwidth}{0.1pt}
\end{figure*}
Here, similar to Definition~\ref{Def:3}, we first introduce a formal definition of the interconnected A-SHS.

\begin{definition}\label{interconnected A-SHS}
	Given $N\in\mathbb N^+$ A-SHSs {$\mathcal{A}_i(\Phi_i)=(\mathbb X_i,\mathbb U_i,\mathscr U_i,\mathds W_i,\mathscr W_i,\sigma_i,\rho_i,\varsigma_i,\mathbb F_i,\mathbb Y_i,\mathbb H_i)$}, with their disturbances partitioned similar to~\eqref{Eq:31}, satisfying interconnection constraint
	\begin{align}\notag
		\forall i,j\in \{1,\dots,N\},i\neq j\!: ~~ w_{ij}=x_{j}, ~~ {\mathbb Y_{j}\subseteq \mathds W_{ij}},
	\end{align}	
	where $\mathds W_i\coloneq\prod_{j\neq i}\mathds W_{ij}$, the interconnection of  $\mathcal{A}_i(\Phi_i)$, $i\in \{1,\ldots,N\}$, forms $\mathcal{A}(\Phi) = (\mathbb X,\mathbb U,\sigma,\rho,\varsigma, \mathbb F,\mathbb Y,\mathbb H)$, denoted by
	$\mathcal{I}(\mathcal{A}_1(\Phi_1),\ldots,\mathcal{A}_N(\Phi_N))$. The interconnected A-SHS $\mathcal A(\Phi)$ is endowed with $\mathbb{X}\!\coloneq\!\prod_{i=1}^{N}\mathbb{X}_i$, $\mathbb U = \prod_{i = 1}^N \mathbb U_i$, $\sigma\!\coloneq\!\mathsf{blkdiag}(\sigma_1(x_1),\ldots,\sigma_N(x_N))$, $\rho\coloneq\mathsf{blkdiag}(\rho_1(x_1),\ldots,\rho_N(x_N))$, $ \varsigma\coloneq[\varsigma_1;\dots;\varsigma_N]$,  $\mathbb{Y}\coloneq\prod_{i=1}^{N} \mathbb{Y}_{i}$, and ${\mathbb{H}:\mathbb X\to\mathbb Y}$, where {$\mathbb H(x,\vartheta) = x$}. Additionally, the map $\mathbb{F}=[\mathbb{F}_{1}; \dots;\mathbb{F}_{N}]$  is defined by {$(x',\vartheta')\!\in \!\mathbb F((x,\vartheta),\nu,\sigma,\rho,\varsigma)$} if and only if, for any $i\in\{1,\dots,N\}$:
	\begin{enumerate}[label=(\roman*)]
		\item Flow case: $0\leq \vartheta_i\leq \varepsilon_{2_i}-1$, $x_i'=  x_{x_i\nu_iw_i}(\tau^-)$, and $\vartheta_i'=\vartheta_i+1$;
		\item Jump case: $\varepsilon_{1_i}\leq \vartheta_i\leq \varepsilon_{2_i}$,  {$x_i'= f_{2_i}(x_i,\nu_i(0),w_i(0),$ $\varsigma_i(0))$}, and $\vartheta_i'=0$; 
	\end{enumerate}
	where $x= [x_1;\dots;x_N]$, $\vartheta= [\vartheta_1;\dots;\vartheta_N]$, $ \varepsilon_1=[\varepsilon_{1_1};\dots;\varepsilon_{1_N}]$, and $\varepsilon_2=[\varepsilon_{2_1};\dots;\varepsilon_{2_N}]$.
\end{definition}

 In the next theorem, we propose the compositionality result of the work, using which one can construct an A-CBC for the interconnected A-SHS based on A-CSBCs of individual subsystems.  To do so, we first define $\Theta\coloneq\mathsf{diag}(\theta_1,\ldots,\theta_N)$ with  $\theta_i = 1 - \gamma_i$, and  $\Psi\coloneq\{\psi_{ij}\}$ with $\psi_{ij} = \frac{\varphi_i}{\kappa_j}$, where $\psi_{ii}=0$, $\forall i\in\{1,\cdots,N\}$. {Note that if there is no connection from $\mathcal{A}_j(\Phi_j)$ to $\mathcal{A}_i(\Phi_i)$, then $w_{ij} \equiv 0$, which, in turn, implies that $\psi_{ij} = 0$.}
 
\begin{theorem}\label{Thm:2}
	Consider an interconnected A-SHS $\mathcal{I}(\mathcal{A}_1(\Phi_1),\ldots,\mathcal{A}_N(\Phi_N))$ induced by A-SHSs~$\mathcal{A}_i(\Phi_i)$, $i\in\{1,\dots,N\}$. Suppose that each $\mathcal{A}_i(\Phi_i)$ admits an A-CSBC $\mathcal B_i$, as defined in Definition~\ref{Def_1a1}. For a vector {$\xi = [\xi_1; \ldots; \xi_N]$}, with $\xi_i >0$, and a scalar value $\Gamma \in (-1,0)$, where $\max_{1\leq i\leq N} \Gamma_i < \Gamma < 0$, if
	\begin{subequations}\label{eq:tmp}
		\begin{align}\label{Eq:21}
			\xi^\top(-\Theta+\Psi)&\coloneq \big[\xi_1\Gamma_1;\dots;\xi_N\Gamma_N\big]^\top < 0,\\\label{Eq:28}
			\sum_{i=1}^N\xi_i \beta_i &> \sum_{i=1}^N\xi_i \mu_i,
		\end{align}
	\end{subequations}
	then 
	\begin{align}\label{Overall B}
		\mathcal B(x,\vartheta) \coloneq \sum_{i=1}^N\xi_i\mathcal B_i(x_i,\vartheta_i)
	\end{align}
	is an A-CBC for the interconnected A-SHS with the initial and unsafe sets $\mathbb X_0\coloneq\prod_{i=1}^{N}\mathbb X_{0_i}$ and $\mathbb X_u\coloneq\prod_{i=1}^{N}\mathbb X_{u_i}$, where
	$$
		\beta \coloneq \sum_{i=1}^{N}\xi_i\beta_i, \; \mu \coloneq \sum_{i=1}^{N} \xi_i\mu_i, \; \eta \coloneq \sum_{i=1}^{N}\xi_i\eta_i,\; \gamma\coloneq 1 + \Gamma.
	$$
\end{theorem}

{\bf Proof.} 
We first show that conditions~\eqref{Eq_2a1} and~\eqref{Eq_2a2} hold. For any $(x,\vartheta) \in \mathbb{X}_0 = \prod_{i=1}^{N} \mathbb{X}_{0_i} $ and from \eqref{Eq_2a11}, one has
$$
	\mathcal B(x,\vartheta) \overset{\eqref{Overall B}}{\vphantom{\leq}=\vphantom{\leq}} \sum_{i=1}^N\xi_i\mathcal B_{i}(x_{i},\vartheta_{i})\overset{\eqref{Eq_2a11}}{\leq}\sum_{i=1}^N\xi_i\mu_{i} = \mu.
$$ 
Similarly, for any $(x,\vartheta) \in \mathbb{X}_{u} = \prod_{i=1}^{N} \mathbb{X}_{u_i} $, and from \eqref{Eq_2a21}, we have
$$
	\mathcal B(x,\vartheta) \overset{\eqref{Overall B}}{\vphantom{\leq}=\vphantom{\leq}} \sum_{i=1}^N\xi_i\mathcal B_{i}(x_{i},\vartheta_{i})\overset{\eqref{Eq_2a21}}{\geq}\sum_{i=1}^N\xi_i\beta_i = \beta,
$$
implying that conditions~\eqref{Eq_2a1} and~\eqref{Eq_2a2} are satisfied with $\mu = \sum_{i=1}^N\xi_i\mu_i$ and $\beta = \sum_{i=1}^N\xi_i\beta_i$. Note that $\beta > \mu $ according to~\eqref{Eq:28}. 

Now, we proceed with showing that condition~\eqref{Eq_3a} holds as well. By leveraging conditions~\eqref{Eq_2a22} and \eqref{Eq:21}, one can acquire the chain of inequalities in \eqref{Eq:23}. By defining $\eta\coloneq\sum_{i=1}^N\xi_i\eta_i$ and
$$
	\gamma s \coloneq \max\Big\{s + \xi^\top(-\Theta+\Psi)\tilde{\mathcal B}(x,\vartheta)\,\big|\, \xi^\top\tilde {\mathcal B}(x,\vartheta)=s\Big\},
$$
with $\tilde {\mathcal B}(x,\vartheta)=\big[\mathcal B_1(x_1,\vartheta_1);\ldots;\mathcal B_N(x_N,\vartheta_N)\big]$, condition \eqref{Eq_3a} is also met. The only remaining part is to show that $ \gamma \coloneq 1 + \Gamma$ with $0<\gamma<1$. Given that $\xi^\top(-\Theta+\Psi) \coloneq \big[\xi_1\Gamma_1;\dots;\xi_N\Gamma_N\big]^\top < 0$, and  $\max_{1\leq i\leq N} \Gamma_i < \Gamma < 0$ with $\Gamma\in (-1,0)$, we have
\begin{align*}
	\gamma s &= s + \xi^\top(-\Theta+\Psi)\tilde {\mathcal B}(x,\vartheta)\\
	& = s \! + \! \big[\xi_1\Gamma_1;\dots;\xi_N\Gamma_N\!\big]^{\!\!\top}\!\big[\!\mathcal B_1(x_1,\vartheta_1);\ldots;\mathcal B_N(x_N,\vartheta_N)\!\big]\\
	&= s \! + \! \xi_1\Gamma_1\mathcal B_1(x_1,\vartheta_1) + \dots + \xi_N\Gamma_N\mathcal B_N(x_N,\vartheta_N)\\
	& \leq s \! + \! \Gamma\big(\xi_1\mathcal B_1(x_1,\vartheta_1) + \dots + \xi_N\mathcal B_N(x_N,\vartheta_N)\big) \\
	&= s \! + \! \Gamma s = (1 + \Gamma) s.
\end{align*}
Therefore, $\gamma s \leq (1 + \Gamma) s$, and consequently $\gamma \leq1 + \Gamma$. Since $\Gamma \in (-1,0)$, then $0 <\gamma \coloneq 1 + \Gamma <1$. Hence, $\mathcal B$ is an A-CBC for $\mathcal{I}(\mathcal{A}_1(\Phi_1),\ldots,\mathcal{A}_N(\Phi_N))$, which concludes the proof. $\hfill\blacksquare$
\begin{remark}
	As the interconnection topology becomes denser, each subsystem is influenced by more neighbors, leading to a stronger aggregate disturbance $w_i$. To account for this in the analysis, the associated disturbance gain $\varphi_i$ may also increase, reflecting higher sensitivity to inter-subsystem coupling. Consequently, the compositional condition~\eqref{Eq:21}, which relies on bounding this influence, becomes more difficult to satisfy, consistent with small-gain reasoning in stability analysis; e.g., see~\citep{dashkovskiy2007iss}.
\end{remark}
As demonstrated in Theorem \ref{Thm:2}, under the compositional conditions in~\eqref{eq:tmp}, an A-CBC for the interconnected A-SHS can be compositionally constructed based on A-CSBCs of its individual subsystems. In the next section, we provide the sufficient conditions for constructing the A-CSBC for the augmented SHS.

\section{Construction of A-CSBC}\label{sec:constrcution_finite}

In this section, we establish conditions on the SHS $\Phi_i$ that allow for the construction of an A-CSBC for $\mathcal{A}_i(\Phi_i)$. To do so, we first define the notion of control barrier certificates over the original $\Phi_i$~\citep{nejati2020compositional, anand2022small}. This intermediate formulation serves as a basis for systematically extending the certificates to the augmented setting to capture hybrid timing behavior.

\begin{definition}\label{cbc}
	Given an SHS $\Phi_i$ with initial and unsafe sets $X_{0_i}, X_{u_i} \subset X_i$, a twice continuously differentiable function $\bar{\mathcal B}_i:X_i \rightarrow \mathbb{R}_{0}^+$ is called a control barrier certificate (CBC) for $\Phi_i$ if there exist constants
	$\gamma_{1_i}\in\R,\bar\kappa_i,\gamma_{2_i}\in\R^+$, $\bar\mu_i,\bar\beta_i,\bar\varphi_{1_i},\bar\varphi_{2_i},\bar\eta_{1_i}, \bar\eta_{2_i}\in\R_{0}^+$, such that
	\begin{subequations}
		\begin{align}\label{subsys21}
			&\bar{\mathcal B}_i(x_i) \geq \bar\kappa_i\Vert x_i\Vert^2,\quad\quad~~~~~\!\forall x_i \in X_{i},\\\label{subsys2}
			&\bar{\mathcal B}_i(x_i) \leq \bar\mu_i,\quad\quad\quad\quad\quad\quad\!\!\forall x_i \in X_{0_i},\\\label{subsys3}
			&\bar{\mathcal B}_i(x_i) \geq \bar\beta_i, \quad\quad\quad\quad\quad\quad\!\forall x_i \in X_{u_i}, 
		\end{align} 
			$\bullet$ $\forall x_i\in X_i$, $\exists \nu_i\in U_i$ such that $\forall w_i\in W_i$,
		\begin{align}\label{Eq:7}
			&\mathcal{L}\bar{\mathcal B}_i(x_i)\leq -\gamma_{1_i} \bar{\mathcal B}_i(x_i) + \bar\varphi_{1_i}\Vert w_i\Vert^2 + \bar\eta_{1_i},
		\end{align}
		{where $\mathcal{L}\bar{\mathcal B}_i(x_i)$ is the infinitesimal generator~\citep{oksendal2013stochastic}  acting on $\bar{\mathcal B}_i(x_i)$, defined as
		\begin{align}\notag
			\mathcal{L}\bar{\mathcal B}_i(x_i) &= \frac{\partial \bar{\mathcal B}_i(x_i)}{\partial x_i} \!f_{1_i}\!(x_i,\!\nu_i,\!w_i)\!\\
   \notag &~~~+ \frac{1}{2}\!\textsf{Tr}~\!(\sigma_i(x_i)\sigma_i(x_i)\!^\top\frac{\partial^2{\bar{\mathcal B}_i(x_i)}}{\partial x_{q_i} \partial x_{q_i'}})\\\notag
			& ~~~+ \sum_{j=1}^{\textsf r_i}\lambda_{j_i} (\,\bar{\mathcal B}_i(x_i+\rho_i(x_i)\textsf e_{j_i}^{\textsf r_i})- \bar{\mathcal B}_i(x_i)),
		\end{align}
		where $q_i,q_i'\in\{1_i,\ldots,n_i\}$, $\lambda_{j{_i}}$ denotes the rates of Poisson processes, and $\textsf e_{j_{{i}}}^{\textsf{r}_{{i}}}$ is an $\textsf{r}_{{i}}$-dimensional vector with $1$ on the $j$-th entry and $0$ elsewhere;}

			$\bullet$ $\forall x_i\in X_i$, $\exists \nu_i\in U_i$ such that $\forall w_i\in W_i$,
		\begin{align}
			\EE\Big[\bar{\mathcal B}_i(f_{2_i}(x_i,\!\nu_i,&w_i,{\varsigma_i})\!) \big| x_i,\!\nu_i,\!w_i\Big]\notag\\
			&\leq\! \gamma_{2_i}\bar{\mathcal B}_i(x_i) \!+\! \bar\varphi_{2_i}\!\Vert w_i\Vert^2\!+\! \bar\eta_{2_i},\label{csbceq}
		\end{align}
		which captures the expected evolution of the certificate under one discrete jump transition.
	\end{subequations}
\end{definition}

\begin{remark}
	 Definition~\ref{cbc} formulates CBCs tailored to SHSs that exhibit both continuous-time and discrete-time (reset) dynamics, as defined in Definition~\ref{Def:SHS}. In this setting, {for instance in the case that $\gamma_{1_i}>0$ and $0<\gamma_{2_i}<1$,} condition~\eqref{Eq:7} ensures the desired decrease of the certificate along the continuous-time evolution, using the infinitesimal generator commonly employed in continuous-time stochastic analysis~\citep{nejati2020compositional}. In contrast, condition~\eqref{csbceq} governs the discrete-time updates and captures the expected decrease across discrete transitions, consistent with the discrete-time stochastic setting discussed in~\citep{anand2022small}. Together, these conditions jointly ensure the safety of the SHS $\Phi_i$ under both modes of evolution.
\end{remark}
Before presenting the main result of this section, we introduce the following lemma, which quantifies an upper bound on the evolution of the CBC $\bar{\mathcal B}_i$. We consider consecutive time instances $t_k = k\tau$, $k \in \mathbb{N}$, where $\tau > 0$ is the sampling interval.

\begin{lemma}\label{Lemma:1}
	Given an SHS $\Phi_i=(X_i, U_i,\mathcal{U}_i,W_i,\mathcal{W}_i,\sigma_i,$ $\rho_i, f_{1_i},\varsigma_i,f_{2_i})$, suppose~\eqref{Eq:7} holds. Then,
	for any $x_i\in X_i$, {there exists $\nu_i\in \mathcal{U}_{\tau_i}$}, such that for any ${w_i\in \mathcal W_i}$, and for any two consecutive time instances $(t_{k},t_{k+1})$, one has  
	\begin{align}\notag
		&\EE \Big[\bar{\mathcal B}_i(x_{x_i\nu_iw_i}(t_{k+1}^-))\,\big|\,x_i,\nu_i,w_i\Big]\\\label{Eq:11}
		&\leq  e^{-\gamma_{1_i}(t_{k+1}-t_{k})}\bar{\mathcal B}_i(x_{x_i\nu_iw_i}(t_{k})\!){+ \Upsilon_i,}
	\end{align}
	{where
	\begin{align}\label{eq:cons}
		\Upsilon_i = \begin{cases}
			e^{-\gamma_{1_i}(t_{k+1}-t_{k})} (t_{k+1} \!-\! t_k) c_i, &\text{if }~ \gamma_{1_i} \leq 0,\\
			{} & {}\\
			\dfrac{1-e^{-\gamma_{1_i}(t_{k+1}-t_{k})}}{\gamma_{1_i}}c_i, & \text{if }~  \gamma_{1_i} > 0,
		\end{cases}
	\end{align}
	with $c_i\coloneq \bar\varphi_{1_i}\Vert w_i\Vert_\infty^2+ \bar\eta_{1_i}$.}
\end{lemma}
{\bf Proof.} From Dynkin's formula~\citep{dynkin1965markov}, we have
\begin{align}
	&\EE \Big[\bar{\mathcal B}_i(x_{x_i\nu_iw_i}(t_{k+1}^-))\,\big|\,x_i,\nu_i,w_i\Big] \notag\\
	&  = \bar{\mathcal B}_i(x_{x_i\nu_iw_i}(t_{k})) \!+\! \EE \Big[\int_{t_k}^{t_{k+1}}\!\!\!\mathcal{L}\bar{\mathcal B}_i(x_{x_i\nu_iw_i}(t))\mathsf{d}t \,\big|\,x_i,\nu_i,w_i\Big]\!. \label{eq:tmp2}
\end{align}
{We note that the value of the integral remains unchanged when using either $t_{k+1}$ or $t_{k+1}^-$ as the upper limit.} From condition~\eqref{Eq:7}, and considering~\eqref{eq:tmp2}, one has
\begin{align}\notag
	&\EE \Big[\bar{\mathcal B}_i(x_{x_i\nu_iw_i}(t_{k+1}^-))\,\big|\,x_i,\nu_i,w_i\Big]\\\notag
	&\leq \bar{\mathcal B}_i(x_{x_i\nu_iw_i}(t_{k}))+\EE \Big[\int_{t_k}^{t_{k+1}}\!\!\!\big(\!-\gamma_{1_i} \bar{\mathcal B}_i(x_{x_i\nu_iw_i}(t)) \\\notag
	&~~~\!+\! \bar\varphi_{1_i}\Vert w_i\Vert^2 + \bar\eta_{1_i}\big)\mathsf{d}t\,\big|\,x_i,\nu_i,w_i\Big]\\\notag
	&\leq \bar{\mathcal B}_i(x_{x_i\nu_iw_i}(t_{k}))+\EE \Big[\int_{t_k}^{t_{k+1}}\!\!\!\big(\!-\gamma_{1_i} \bar{\mathcal B}_i(x_{x_i\nu_iw_i}(t)) \\\notag
	&~~~\!+\!  \underbrace{\bar\varphi_{1_i}\Vert w_i\Vert_\infty^2 \!+ \bar\eta_{1_i}}_{c_i}\big)\mathsf{d}t\,\big|\,x_i,\nu_i,w_i\Big]\notag\\
&=\bar{\mathcal B}_i(x_{x_i\nu_iw_i}(t_{k})) + (t_{k+1}-t_k)c_i\notag\\
	&\hphantom{\leq}+\int_{t_k}^{t_{k+1}}\!\!\!\big(-\gamma_{1_i}\EE \Big[\bar{\mathcal B}_i(x_{x_i\nu_iw_i}(t))\,\big|\,x_i,\nu_i,w_i\Big]\big)\mathsf{d}t.\label{eq:lem1}
\end{align}
{When $\gamma_{1_i}\leq 0$ (equivalently $-\gamma_{1_i}\geq 0$), one can employ Gronwall inequality~\citep{gronwall1919note} and obtain
\begin{align*}
	&\EE \Big[\bar{\mathcal B}_i(x_{x_i\nu_iw_i}(t_{k+1}^-))\,\big|\,x_i,\nu_i,w_i\Big]\\
	&\leq  e^{-\gamma_{1_i}(t_{k+1}-t_{k})}\bar{\mathcal B}_i(x_{x_i\nu_iw_i}(t_{k}))\\
	&\hphantom{\leq}~+ e^{-\gamma_{1_i}(t_{k+1}-t_{k})} (t_{k+1} \!-\! t_k) c_i\\
	&=  e^{-\gamma_{1_i}(t_{k+1}-t_{k})}\bar{\mathcal B}_i(x_{x_i\nu_iw_i}(t_{k}))\\
	&\hphantom{\leq}~+ e^{-\gamma_{1_i}(t_{k+1}-t_{k})} (t_{k+1} \!-\! t_k) (\bar\varphi_{1_i}\Vert w_i\Vert_\infty^2+ \bar\eta_{1_i}).
\end{align*}
On the other hand, if $\gamma_{1_i}>0$ (equivalently $-\gamma_{1_i}< 0$), we aim to obtain a bound using the solution of the integral. For any $\hat t\in[t_k,t_{k+1})$, let us define $\aleph_i(\hat t)\coloneq\EE\!\left[\bar{\mathcal B}_i(x_{x_i\nu_iw_i}(\hat t))\,\middle|\,x_i,\nu_i,w_i\right]$. From~\eqref{eq:lem1}, we have
\begin{align}\label{eq:int}
	\aleph_i(\hat t)\leq \aleph_i(t_k)+\int_{t_k}^{\hat t}\!\big(-\!\gamma_{1_i}\aleph_i(t)+c_i\big)\mathsf d t, \, \forall \hat t\in [t_k,t_{k+1}).
\end{align}
Note that, since $x_{x_i\nu_iw_i}(t_k)=x_i$, one has $\aleph_i(t_k)=\EE\left[\bar{\mathcal B}_i(x_{x_i\nu_iw_i}(t_k))\middle|x_i,\nu_i,w_i\right]=\bar{\mathcal B}_i(x_{x_i\nu_iw_i}(t_k))$.
Since by Dynkin’s formula in~\eqref{eq:tmp2}, the map $\hat t \mapsto \aleph_i(\hat t)$ is absolutely continuous on $[t_k, t_{k+1})$, differentiating both sides of  \eqref{eq:int} with respect to $\hat t$ yields
\begin{align}\label{eq:int 2}
	\frac{\mathsf d\aleph_i(\hat t)}{\mathsf d \hat t}\leq -\gamma_{1_i}\aleph_i(\hat t) + c_i, \text{ for almost every } \hat t\in[t_k,t_{k+1}).
\end{align}
Now, let $\Lambda_i(\hat t)\coloneq\aleph_i(\hat t)-\frac{c_i}{\gamma_{1_i}}$, whose derivative with respect to $\hat t$ can be obtained as
\begin{align}
	\frac{\mathsf d\Lambda_i(\hat t)}{\mathsf d \hat t} = \frac{\mathsf d\aleph_i(\hat t)}{\mathsf d \hat t}&\overset{\eqref{eq:int 2}}{\leq} -\gamma_{1_i}\aleph_i(\hat t) + c_i \notag\\
	&\overset{\hphantom{\eqref{eq:int 2}}}{=} -\gamma_{1_i}\big(\underbrace{\aleph_i(\hat t) - \frac{c_i}{\gamma_{1_i}}}_{\Lambda_i(\hat t)}\big)\notag\\
	&\overset{\hphantom{\eqref{eq:int 2}}}{=}-\gamma_{1_i}\Lambda_i(\hat t).\label{eq:int 3}
\end{align}
Multiplying \eqref{eq:int 3} by $e^{\gamma_{1_i}(\hat t - t_k)}$ and moving all terms to the left-hand side of the inequality yields
\begin{align*}
	&e^{\gamma_{1_i}(\hat t - t_k)}\frac{\mathsf d\Lambda_i(\hat t)}{\mathsf d \hat t}+\gamma_{1_i}e^{\gamma_{1_i}(\hat t - t_k)}\Lambda_i(\hat t)\\
	&=\frac{\mathsf d}{\mathsf d \hat t}\big(e^{\gamma_{1_i}(\hat t - t_k)}\Lambda_i(\hat t)\big)\leq 0,
\end{align*}
which implies $e^{\gamma_{1_i}(\hat t - t_k)}\Lambda_i(\hat t)$ is non-increasing.
Hence, for any $\hat t\in[t_k, t_{k+1})$, we have
\begin{align*}
	e^{\gamma_{1_i}(\hat t - t_k)}\Lambda_i(\hat t)\leq \Lambda_i(t_{k}),
\end{align*}
or equivalently,
\begin{align}\label{eq:int 4}
	\Lambda_i(\hat t)\leq e^{-\gamma_{1_i}(\hat t - t_k)}\Lambda_i(t_{k}).
\end{align}
Substituting back $\Lambda_i(\hat{t})=\aleph_i(\hat{t})-\frac{c_i}{\gamma_{1_i}}$ in \eqref{eq:int 4}, we obtain
$$
\aleph_i(\hat{t})-\frac{c_i}{\gamma_{1_i}} \leq(\aleph_i(t_k)-\frac{c_i}{\gamma_{1_i}}) e^{-\gamma_{1_i}(\hat{t}-t_k)},
$$
and consequently,
\begin{align}\label{eq:int 5}
	\aleph_i(\hat{t}) \leq \aleph_i(t_k) e^{-\gamma_{1_i}\!(\hat{t}-t_k)}+\frac{1-e^{-\gamma_{1_i}\!(\hat{t}-t_k)}}{\gamma_{1_i}}c_i.
\end{align}
Substituting the definition of $\aleph_i(\hat t)$ in~\eqref{eq:int 5}, by taking $\hat t=t_{k+1}^-$, and recalling that the upper limit of the integral as $t_{k+1}$ or $t_{k+1}^-$ does not change its value, one gets
\begin{align*}
	&\EE\!\Big[\bar{\mathcal B}_i(x_{x_i\nu_iw_i}(t_{k+1}^-))\,\big|\,x_i,\nu_i,w_i\Big]\\
	&\leq e^{-\gamma_{1_i}\!(t_{k+1}-t_k)}\bar{\mathcal B}_i(x_{x_i\nu_iw_i}(t_k))\!+\!\frac{\big(1\!-\! e^{-\gamma_{1_i}\!(t_{k+1}-t_k)}\big)}{\gamma_{1_i}}c_i\\
	&= e^{-\gamma_{1_i}\!(t_{k+1}-t_k)}\bar{\mathcal B}_i(x_{x_i\nu_iw_i}(t_k))\\
	&\hphantom{\leq}+\!\frac{\big(1\!-\! e^{-\gamma_{1_i}\!(t_{k+1}-t_k)}\big)}{\gamma_{1_i}}(\bar\varphi_{1_i}\Vert w_i\Vert_\infty^2+ \bar\eta_{1_i}),
\end{align*}
which concludes the proof. $\hfill\blacksquare$}

Under Definition~\ref{cbc} and Lemma~\ref{Lemma:1}, we propose the next theorem, as the main result of the section, to construct an A-CSBC for $\mathcal{A}_i(\Phi_i)$ based on the CBC $\bar{\mathcal B}_i$.

\begin{theorem}\label{Thm:Main}
	Given an SHS $\Phi_i=(X_i, U_i,\mathcal{U}_i,W_i,\mathcal{W}_i,$ $\sigma_i,\rho_i, f_{1_i},\varsigma_i,f_{2_i})$ with its associated A-SHS {$\mathcal{A}_i(\Phi_i)=(\mathbb X_i,\mathbb U_i,\mathscr U_i,\mathds W_i,\mathscr W_i,\sigma_i,\rho_i,\varsigma_i,\mathbb F_i,\mathbb Y_i,\mathbb H_i)$}, let $\bar{\mathcal B}_i$ be a CBC for  $\Phi_i$. If
	\begin{align}\label{Eq:12}
		\ln(\gamma_{2_i})-\gamma_{1_i}\tau \vartheta_i<0, \quad \forall \vartheta_i\in\{\varepsilon_{1_i},\dots,\varepsilon_{2_i}\},
	\end{align}
	then 
	\begin{align}\label{Eq:13}
		\mathcal{B}_i(x_i,\vartheta_i)=\varpi_i(\vartheta_i)~\!
		\bar{\mathcal B}_i(x_i)
	\end{align}
	is an A-CSBC for $\mathcal{A}_i(\Phi_i)$, with
	\begin{align}\label{beta}
    \varpi_i(\vartheta_i)\coloneq\begin{cases}
        1, & \text{if }~\gamma_{1_i}>0~\&~0<\gamma_{2_i}<1,\\
        e^{\gamma_{1_i}\tau \alpha_{1_i} \vartheta_i}, & \text{if }~\gamma_{1_i}>0 ~\&~ \gamma_{2_i}\geq 1,\\
        \gamma_{2_i}^{\frac{\vartheta_i}{\alpha_{2_i}}}, & \text{if }~\gamma_{1_i}\leq 0 ~\&~ 0<\gamma_{2_i}<1,
    \end{cases}
	\end{align}
	for some $0<\alpha_{1_i}<1$ and $\alpha_{2_i}>\varepsilon_{2_i}$. 
\end{theorem}

{\bf Proof.} By employing \eqref{subsys21}, for any $(x_i,\vartheta_i)\in \mathbb X_{i}$, one has
\begin{align*}
	\mathcal{B}_i(x_i,\vartheta_i)\overset{\eqref{Eq:13}}{\vphantom{\leq}=\vphantom{\leq}}\varpi_i(\vartheta_i)~\!
	\bar{\mathcal B}_i(x_i) \overset{\eqref{subsys21}}{\geq} \varpi_i(\vartheta_i)\bar\kappa_i\Vert x_i\Vert^2.
\end{align*}
Then, condition~\eqref{Eq_2a22} holds with $$\kappa_i=\big(\min_{\vartheta_i\in\{0,\ldots,\varepsilon_{2_i}\}}\varpi_i(\vartheta_i)\big)\bar{\kappa}_i.$$ By leveraging \eqref{subsys2}, for any $(x_i,\vartheta_i)\in \mathbb X_{0_i}$, one has
\begin{align*}
	\mathcal{B}_i(x_i,\vartheta_i)\overset{\eqref{Eq:13}}{\vphantom{\leq}=\vphantom{\leq}}\varpi_i(\vartheta_i)~\!
	\bar{\mathcal B}_i(x_i) \overset{\eqref{subsys2}}{\leq} \varpi_i(\vartheta_i)~\! \bar \mu_i,
\end{align*}
and similarly, by utilizing~\eqref{subsys3} for any $(x_i,\vartheta_i)\in \mathbb X_{u_i}$, one has
\begin{align*}
	\mathcal{B}_i(x_i,\vartheta_i)\overset{\eqref{Eq:13}}{\vphantom{\leq}=\vphantom{\leq}}\varpi_i(\vartheta_i)~\!
	\bar{\mathcal B}_i(x_i) \overset{\eqref{subsys3}}{\geq} \varpi_i(\vartheta_i)~\! \bar \beta_i,
\end{align*}
satisfying conditions \eqref{Eq_2a21} and \eqref{Eq_2a11} with 
\begin{align*}
\beta_i=\big(\min_{\vartheta_i\in\{0,\ldots,\varepsilon_{2_i}\}}\varpi_i(\vartheta_i)\big)\bar{\beta}_i,
\end{align*}
and $\mu_i = \varpi_i(0)\bar\mu_i = \bar\mu_i$, since $\vartheta_{i}=0$ in $\mathbb X_{0_i}$, and consequently, $\varpi_i(0)=1$ for all bounds.

Now, we proceed with showing condition~\eqref{Eq_3a1} as well.
From \eqref{Eq:11} {and \eqref{eq:cons}} with $t_{k+1}=\tau, t_{k}=0$, one gets 
\begin{align}
	&\EE \Big[\bar{\mathcal B}_i(x_{x_i\nu_iw_i}(\tau^-))\,\big|\,x_i,\nu_i,w_i\Big] \notag\\
	&\leq  \!e^{-\gamma_{1_i}\!\tau}\bar{\mathcal B}_i(x_{x_i\nu_iw_i}(0)) \!+\! {\Upsilon_i^\tau} \!=\! e^{-\gamma_{1_i}\!\tau}\bar{\mathcal B}_i(x_i) \!+\! {\Upsilon_i^\tau},\label{eq:cons tau}
\end{align}
where
\begin{align*}
	\Upsilon_i^\tau = \begin{cases}
		e^{-\gamma_{1_i}\tau} \tau (\bar\varphi_{1_i}\Vert w_i\Vert_\infty^2+ \bar\eta_{1_i}), & \text{if }~ \gamma_{1_i} \leq 0,\\
		{} & {}\\
		\dfrac{\big(1-e^{-\gamma_{1_i}\tau}\big)}{\gamma_{1_i}}(\bar\varphi_{1_i}\Vert w_i\Vert_\infty^2+ \bar\eta_{1_i}), & \text{if }~  \gamma_{1_i} > 0.
	\end{cases}
\end{align*}
In order to show that $\mathcal{B}_i$, defined in \eqref{Eq:13}, fulfills condition~\eqref{Eq_3a1}, we consider the two cases in Definition \ref{Def:A-SHS} and the three bounds for different values of $\gamma_{1_i}$ and $\gamma_{2_i}$ in~\eqref{beta} as follows:

    $\bullet$ \textbf{Bound 1} ($\gamma_{1_i}>0~\&~0<\gamma_{2_i}<1$):
		
        (i) \textbf{Flow case} $(0\leq \vartheta_i\leq \varepsilon_{2_i}-1, \vartheta_i'= \vartheta_i+1)$:
		\begin{align*}
			&\EE \Big[\mathcal B_i(x_i',\vartheta_i')\,\big|\,x_i,\!\nu_i,\!w_i,\! \vartheta_i\Big]\!\!\overset{\eqref{Eq:13}}{=}\!\EE \Big[ \bar{\mathcal B}_i(x_i')\,\big|\,x_i,\! \nu_i,\!w_i \Big]\\
			&\overset{\eqref{eq:cons tau}}{\leq} \!e^{-\gamma_{1_i}\!\tau}\bar{\mathcal B}_i(x_i) \!+\! \tfrac{(1-e^{-\gamma_{1_i}\!\tau})}{\gamma_{1_i}}\bar\varphi_{1_i}\Vert w_i\Vert_\infty^2\!+\! \tfrac{(1-e^{-\gamma_{1_i}\!\tau})}{\gamma_{1_i}} \bar\eta_{1_i}\\
			&\overset{\eqref{Eq:13}}{=} \!e^{-\gamma_{1_i}\!\tau}\mathcal B_i(x_i,\!\vartheta_i)\!+\! \tfrac{(1-e^{-\gamma_{1_i}\!\tau})}{\gamma_{1_i}}\bar\varphi_{1_i}\Vert w_i\Vert_\infty^2\\
			&\hphantom{\overset{\hphantom{\eqref{eq:cons tau}}}{\leq}}+\! \tfrac{(1-e^{-\gamma_{1_i}\!\tau})}{\gamma_{1_i}} \bar\eta_{1_i}\\
			&\overset{\hphantom{\eqref{eq:cons tau}}}{\leq}\!e^{-\gamma_{1_i}\!\tau}\mathcal B_i(x_i,\!\vartheta_i) \!+\! \tfrac{(1-e^{-\gamma_{1_i}\!\tau})}{\gamma_{1_i}} \bar\varphi_{1_i}(\Vert w_i\Vert_\infty^2 \!+\! \Vert w_i\Vert^2) \\
			&\hphantom{\overset{\hphantom{\eqref{eq:cons tau}}}{\leq}}+ \tfrac{(1-e^{-\gamma_{1_i}\!\tau})}{\gamma_{1_i}} \bar\eta_{1_i}\\
			&\overset{\hphantom{\eqref{Eq:13}}}{=} \!e^{-\gamma_{1_i}\!\tau}\mathcal B_i(x_i,\!\vartheta_i) \!+\! \tfrac{(1-e^{-\gamma_{1_i}\!\tau})}{\gamma_{1_i}} \bar\varphi_{1_i} \Vert w_i\Vert^2\\
			&\hphantom{\overset{\hphantom{\eqref{eq:cons tau}}}{\leq}}+ \tfrac{(1-e^{-\gamma_{1_i}\!\tau})}{\gamma_{1_i}} \bar\varphi_{1_i} \Vert w_i\Vert_\infty^2 + \tfrac{(1-e^{-\gamma_{1_i}\!\tau})}{\gamma_{1_i}} \bar\eta_{1_i}.
		\end{align*}
		(ii) \textbf{Jump case} $(\varepsilon_{1_i}\leq \vartheta_i\leq \varepsilon_{2_i}, \vartheta_i' = 0)$:
		\begin{align*}
			\EE &\Big[\mathcal B_i(x_i',\vartheta_i')\,\big|\,x_i,\!\nu_i,\!w_i,\! \vartheta_i\Big]\!\overset{\eqref{Eq:13}}{=}\!\EE \Big[ \bar{\mathcal B}_i(x_i')\,\big|\,x_i, \!\nu_i,\!w_i \Big]\\
			&\overset{\eqref{csbceq}}{\leq} \gamma_{2_i}\bar{\mathcal B}_i(x_i) \!+\! \bar\varphi_{2_i}\Vert w_i\Vert^2 \!+\! \bar\eta_{2_i}\\
			&\overset{\eqref{Eq:13}}{=}\gamma_{2_i}\mathcal B_i(x_i,\vartheta_i) \!+\! \bar\varphi_{2_i}\Vert w_i\Vert^2 + \bar\eta_{2_i}.
		\end{align*}
    By defining
    \begin{align*}
        \gamma_i&=\max\Big\{e^{-\gamma_{1_i}\tau},\;\gamma_{2_i}\Big\}, {\varphi_i=\max\Big\{\tfrac{(1-e^{-\gamma_{1_i}\!\tau})}{\gamma_{1_i}} \bar\varphi_{1_i},\bar\varphi_{2_i}\Big\}},\\
        {\eta_i}&{=\max\Big\{\tfrac{(1-e^{-\gamma_{1_i}\!\tau})}{\gamma_{1_i}} \bar\varphi_{1_i} \Vert w_i\Vert_\infty^2 + \tfrac{(1-e^{-\gamma_{1_i}\!\tau})}{\gamma_{1_i}} \bar\eta_{1_i}, \bar\eta_{2_i}\Big\},}
    \end{align*}
    we have
		\begin{align*}
			\EE \Big[\mathcal B_i(x_i',\!\vartheta_i') \big|x_i,\!\nu_i,\!w_i,\!\vartheta_i\Big]&\leq \gamma_i \mathcal B_i(x_i,\!\vartheta_i)+\varphi_{i}\Vert w_i\Vert^2\!+\eta_i.
		\end{align*}
	{It is noteworthy that since $\eta_i$ is constant, we use a uniform bound on $\Vert w_i\Vert_\infty^2$ in its derivation by computing $\sup_{w_i\in\mathcal W_i}\Vert w_i\Vert_\infty^2<\infty$. The same argument applies to the rest of the proof as well.}
	
    $\bullet$ \textbf{Bound 2} ($\gamma_{1_i}>0~\&~\gamma_{2_i}\geq1$):
		
        (i) \textbf{Flow case} $(0\leq \vartheta_i\leq \varepsilon_{2_i}-1, \vartheta_i'= \vartheta_i+1)$:
		\begin{align*}
			&\EE \Big[\mathcal B_i(x_i',\vartheta_i')\,\big|\,x_i,\nu_i,w_i, \vartheta_i\Big]\\
			&\overset{\eqref{Eq:13}}{=}\!e^{\gamma_{1_i}\tau \alpha_{1_i} \vartheta_i'}\EE \Big[ \bar{\mathcal B}_i(x_i')\,\big|\,x_i, \nu_i,w_i \Big]\\
			&\overset{\hphantom{\eqref{Eq:13}}}{=}\! e^{\gamma_{1_i}\tau \alpha_{1_i}\! (\vartheta_i+1)}\EE \Big[ \bar{\mathcal B}_i(x_i')\,\big|\,x_i, \nu_i,w_i \Big]\\
			&\overset{\eqref{eq:cons tau}}{\leq}\! e^{\gamma_{1_i}\tau \alpha_{1_i}\! (\vartheta_i+1)}\big(e^{-\gamma_{1_i}\tau}\bar{\mathcal B}_i(x_i)\!+\! \tfrac{(1-e^{-\gamma_{1_i}\!\tau})}{\gamma_{1_i}}\bar\varphi_{1_i}\Vert w_i\Vert_\infty^2\\
            &\hphantom{\overset{\eqref{eq:cons tau}}{\leq}}+ \tfrac{(1-e^{-\gamma_{1_i}\!\tau})}{\gamma_{1_i}}\bar\eta_{1_i}\big)\\
			&\overset{\hphantom{\eqref{Eq:13}}}{=}\!e^{-\gamma_{1_i}\tau}\!e^{\gamma_{1_i}\tau \alpha_{1_i}}e^{\gamma_{1_i}\tau \alpha_{1_i} \vartheta_i} \bar{\mathcal B}_i(x_i)\\
			&\hphantom{\overset{\eqref{Eq:13}}{=}} +e^{\gamma_{1_i}\tau \alpha_{1_i}\! (\vartheta_i+1)}\tfrac{(1-e^{-\gamma_{1_i}\!\tau})}{\gamma_{1_i}}\bar\varphi_{1_i}\Vert w_i\Vert_\infty^2 \\
			&\hphantom{\overset{\eqref{Eq:13}}{=}}+ e^{\gamma_{1_i}\tau \alpha_{1_i}\! (\vartheta_i+1)}\tfrac{(1-e^{-\gamma_{1_i}\!\tau})}{\gamma_{1_i}}\bar\eta_{1_i}\\ 
			&\overset{\eqref{Eq:13}}{=}\!e^{-\gamma_{1_i}\tau(1-\alpha_{1_i})}\mathcal B_i(x_i,\vartheta_i)\!\\ &\hphantom{\overset{\eqref{Eq:13}}{=}}+e^{\gamma_{1_i}\tau \alpha_{1_i}\! (\vartheta_i+1)}\tfrac{(1-e^{-\gamma_{1_i}\!\tau})}{\gamma_{1_i}} \bar\varphi_{1_i}\Vert w_i\Vert_\infty^2\\
			&\hphantom{\overset{\eqref{Eq:13}}{=}} + e^{\gamma_{1_i}\tau \alpha_{1_i}\! (\vartheta_i+1)}\tfrac{(1-e^{-\gamma_{1_i}\!\tau})}{\gamma_{1_i}}\bar\eta_{1_i}\\ 
			&\overset{\hphantom{\eqref{eq:cons tau}}}{\leq}\! e^{-\gamma_{1_i}\tau(1-\alpha_{1_i})}\mathcal B_i(x_i,\vartheta_i)\\ &\hphantom{\overset{\eqref{eq:cons tau}}{\leq}}+e^{\gamma_{1_i}\tau \alpha_{1_i} \varepsilon_{2_i}}\tfrac{(1-e^{-\gamma_{1_i}\!\tau})}{\gamma_{1_i}} \bar\varphi_{1_i}\Vert w_i\Vert_\infty^2\\
			&\hphantom{\overset{\eqref{eq:cons tau}}{\leq}}+ e^{\gamma_{1_i}\tau \alpha_{1_i} \varepsilon_{2_i}}\tfrac{(1-e^{-\gamma_{1_i}\!\tau})}{\gamma_{1_i}}\bar\eta_{1_i}\\
            &\overset{\hphantom{\eqref{eq:cons tau}}}{\leq}\! e^{-\gamma_{1_i}\tau(1-\alpha_{1_i})}\mathcal B_i(x_i,\vartheta_i) \\ &\hphantom{\overset{\eqref{eq:cons tau}}{\leq}}+e^{\gamma_{1_i}\tau \alpha_{1_i} \varepsilon_{2_i}}\tfrac{(1-e^{-\gamma_{1_i}\!\tau})}{\gamma_{1_i}} \bar\varphi_{1_i}(\Vert w_i\Vert_\infty^2 + \Vert w_i\Vert^2)\\ &\hphantom{\overset{\eqref{eq:cons tau}}{\leq}}+ e^{\gamma_{1_i}\tau \alpha_{1_i} \varepsilon_{2_i}}\tfrac{(1-e^{-\gamma_{1_i}\!\tau})}{\gamma_{1_i}}\bar\eta_{1_i}\\
            &\overset{\hphantom{\eqref{Eq:13}}}{=}\! e^{-\gamma_{1_i}\tau(1-\alpha_{1_i})}\mathcal B_i(x_i,\vartheta_i)\\
            &\hphantom{\overset{\eqref{Eq:13}}{=}}+e^{\gamma_{1_i}\tau \alpha_{1_i} \varepsilon_{2_i}}\tfrac{(1-e^{-\gamma_{1_i}\!\tau})}{\gamma_{1_i}} \bar\varphi_{1_i}\Vert w_i\Vert^2\\
            &\hphantom{\overset{\eqref{Eq:13}}{=}}+ e^{\gamma_{1_i}\tau \alpha_{1_i} \varepsilon_{2_i}}\tfrac{(1-e^{-\gamma_{1_i}\!\tau})}{\gamma_{1_i}}\bar\varphi_{1_i}\Vert w_i\Vert_\infty^2\\  
			&\hphantom{\overset{\eqref{Eq:13}}{=}}+ e^{\gamma_{1_i}\tau \alpha_{1_i} \varepsilon_{2_i}}\tfrac{(1-e^{-\gamma_{1_i}\!\tau})}{\gamma_{1_i}}\bar\eta_{1_i}.
		\end{align*}
		(ii) \textbf{Jump case} $(\varepsilon_{1_i}\leq \vartheta_i\leq \varepsilon_{2_i}, \vartheta_i' = 0)$:
		\begin{align*}
			\EE &\Big[\mathcal B_i(x_i',\vartheta_i')\,\big|\,x_i,\nu_i,w_i, \vartheta_i\Big]\\
			&\overset{\eqref{Eq:13}}{=}e^{\gamma_{1_i}\tau \alpha_{1_i} \vartheta_i'}\EE \Big[ \bar{\mathcal B}_i(x_i')\,\big|\,x_i, \nu_i,w_i \Big]\\
			&\overset{\hphantom{\eqref{Eq:13}}}{=}\EE \Big[ \bar{\mathcal B}_i(x_i')\,\big|\,x_i, \nu_i,w_i \Big]\\
			& \overset{\eqref{csbceq}}{\leq} \gamma_{2_i}\bar{\mathcal B}_i(x_i) \!+\! \bar\varphi_{2_i}\Vert w_i\Vert^2 \!+\! \bar\eta_{2_i}\\
			&\overset{\hphantom{\eqref{Eq:13}}}{=}\frac{e^{\gamma_{1_i}\tau \alpha_{1_i} \vartheta_i}}{e^{\gamma_{1_i}\tau \alpha_{1_i} \vartheta_i}}\gamma_{2_i}\bar{\mathcal B}_i(x_i) \!+\! \bar\varphi_{2_i}\Vert w_i\Vert^2 \!+\! \bar\eta_{2_i}\\
			&\overset{\eqref{Eq:13}}{\leq}e^{-\gamma_{1_i}\tau \alpha_{1_i} \varepsilon_{1_i}}\gamma_{2_i}\mathcal B_i(x_i,\vartheta_i)\!+\! \bar\varphi_{2_i}\Vert w_i\Vert^2 \!+\! \bar\eta_{2_i}. 
		\end{align*}
      By defining 
       \begin{align*}
        \gamma_i&=\max\{e^{-\gamma_{1_i}\tau(1-\alpha_{1_i})},e^{-\gamma_{1_i}\tau \alpha_{1_i} \varepsilon_{1_i}}\gamma_{2_i}\},\\
        {\varphi_i}&{=\max\{e^{\gamma_{1_i}\tau \alpha_{1_i} \varepsilon_{2_i}}\tfrac{(1-e^{-\gamma_{1_i}\!\tau})}{\gamma_{1_i}} \bar\varphi_{1_i},\bar\varphi_{2_i}\}},\\
      {\eta_i}&{=\max\{e^{\gamma_{1_i}\tau \alpha_{1_i} \varepsilon_{2_i}}\tfrac{(1-e^{-\gamma_{1_i}\!\tau})}{\gamma_{1_i}}\bar\varphi_{1_i}\Vert w_i\Vert_\infty^2}\\
      &{\hphantom{=\max\{}+ e^{\gamma_{1_i}\tau \alpha_{1_i} \varepsilon_{2_i}}\tfrac{(1-e^{-\gamma_{1_i}\!\tau})}{\gamma_{1_i}}\bar\eta_{1_i},\bar\eta_{2_i}\}},
  \end{align*}
   we have 
    \begin{align*}
        \EE \Big[\mathcal B_i(x_i',\vartheta_i')\big|x_i,\!\nu_i,\!w_i,\!\vartheta_i\Big]&\leq \gamma_i \mathcal B_i(x_i,\vartheta_i)\!+\!\varphi_{i}\Vert w_i\Vert^2\!\!+\!\eta_i.
    \end{align*}
    $\bullet$ \textbf{Bound 3} ($\gamma_{1_i}\leq0 ~\&~ 0<\gamma_{2_i}<1$):
	
    (i) \textbf{Flow case} $(0\leq \vartheta_i\leq \varepsilon_{2_i}-1, \vartheta_i'= \vartheta_i+1)$:
  \begin{align*}
         &\EE \Big[\mathcal B_i(x_i',\vartheta_i')\,\big|\,x_i,\nu_i,w_i,\vartheta_i\Big]\!\!\overset{\eqref{Eq:13}}{=}\!\!\gamma_{2_i}^{\frac{\vartheta_i'}{\alpha_{2_i}}}\EE \Big[ \bar{\mathcal B}_i(x_i')\big|x_i, \nu_i,w_i \Big] \\
         &\overset{\hphantom{\eqref{Eq:13}}}{=}\!\!\!\gamma_{2_i}^{\frac{\vartheta_i+1}{\alpha_{2_i}}}\EE \Big[ \bar{\mathcal B}_i(x_i')\,\big|\,x_i, \nu_i,w_i \Big] \\
         &\overset{\eqref{eq:cons tau}}{\leq}\!\!\!\gamma_{2_i}^{\frac{\vartheta_i+1}{\alpha_{2_i}}}\!\Big(\!e^{-\gamma_{1_i}\!\tau}\bar{\mathcal B}_i(x_i) \!+\! e^{-\gamma_{1_i}\!\tau} \tau\bar\varphi_{1_i}\Vert w_i\Vert_\infty^2 \!\!+\! e^{-\gamma_{1_i}\!\tau} \tau\bar\eta_{1_i}\!\Big)\\
        	&\overset{\hphantom{\eqref{Eq:13}}}{=}\!\!e^{-\gamma_{1_i}\tau}\gamma_{2_i}^{\frac{1}{\alpha_{2_i}}}\gamma_{2_i}^{\frac{\vartheta_i}{\alpha_{2_i}}} \bar{\mathcal B}_i(x_i) + \gamma_{2_i}^{\frac{\vartheta_i+1}{\alpha_{2_i}}}e^{-\gamma_{1_i}\tau} \tau\bar\varphi_{1_i}\Vert w_i\Vert_\infty^2\\
            &\hphantom{\overset{\eqref{Eq:13}}{=}}+ \gamma_{2_i}^{\frac{\vartheta_i+1}{\alpha_{2_i}}}\!e^{-\gamma_{1_i}\tau}\tau\bar\eta_{1_i} \\
        	&\overset{\eqref{Eq:13}}{\leq}\!\! e^{-\gamma_{1_i}\tau}\gamma_{2_i}^{\frac{1}{\alpha_{2_i}}} \mathcal B_i(x_i,\vartheta_i)+ \gamma_{2_i}^{\frac{1}{\alpha_{2_i}}}e^{-\gamma_{1_i}\tau} \tau\bar\varphi_{1_i}\Vert w_i\Vert_\infty^2 \\
            &\hphantom{\overset{\eqref{Eq:13}}{\leq}}+ \gamma_{2_i}^{\frac{1}{\alpha_{2_i}}}e^{-\gamma_{1_i}\tau}\tau\bar\eta_{1_i}\\
         &\overset{\hphantom{\eqref{Eq:13}}}{\leq}\!\!e^{-\gamma_{1_i}\tau}\gamma_{2_i}^{\frac{1}{\alpha_{2_i}}} \mathcal B_i(x_i,\vartheta_i) \\
         &\hphantom{\overset{\eqref{Eq:13}}{\leq}}+ \gamma_{2_i}^{\frac{1}{\alpha_{2_i}}}e^{-\gamma_{1_i}\tau} \tau\bar\varphi_{1_i}(\Vert w_i\Vert_\infty^2 + \Vert w_i\Vert^2) \\
         &\hphantom{\overset{\eqref{Eq:13}}{\leq}}+ \gamma_{2_i}^{\frac{1}{\alpha_{2_i}}}e^{-\gamma_{1_i}\tau}\tau\bar\eta_{1_i} \\
         &\overset{\hphantom{\eqref{Eq:13}}}{=}\!\! e^{-\gamma_{1_i}\tau}\gamma_{2_i}^{\frac{1}{\alpha_{2_i}}} \mathcal B_i(x_i,\vartheta_i)\!+\! \gamma_{2_i}^{\frac{1}{\alpha_{2_i}}}e^{-\gamma_{1_i}\tau} \tau\bar\varphi_{1_i}\Vert w_i\Vert^2 \\
         &\hphantom{\overset{\eqref{Eq:13}}{=}} + \gamma_{2_i}^{\frac{1}{\alpha_{2_i}}}e^{-\gamma_{1_i}\tau} \tau\bar\varphi_{1_i}\Vert w_i\Vert_\infty^2 + \gamma_{2_i}^{\frac{1}{\alpha_{2_i}}}e^{-\gamma_{1_i}\tau}\tau\bar\eta_{1_i}.
\end{align*}
		(ii) \textbf{Jump case} $(\varepsilon_{1_i}\leq \vartheta_i\leq \varepsilon_{2_i}, \vartheta_i' = 0)$:
		\begin{align*}
			&\EE \Big[\mathcal B_i(x_i',\vartheta_i')\,\big|\,x_i,\nu_i,w_i, \vartheta_i\Big]\!\!\overset{\eqref{Eq:13}}{=}\!\!\gamma_{2_i}^{\frac{\vartheta_i'}{\alpha_{2_i}}}\EE \Big[ \bar{\mathcal B}_i(x_i')\big|x_i, \nu_i,w_i \Big]\\
			&\overset{\hphantom{\eqref{Eq:13}}}{=}\!\!\EE \Big[ \bar{\mathcal B}_i(x_i')\big|x_i,\! \nu_i,\!w_i \Big]\!\! \overset{\eqref{csbceq}}{\leq}\!\! \gamma_{2_i}\bar{\mathcal B}_i(x_i) \!+\! \bar\varphi_{2_i}\Vert w_i\Vert^2 \!+\! \bar\eta_{2_i}\\
			&\overset{\hphantom{\eqref{Eq:13}}}{=}\!\!\frac{\gamma_{2_i}^{\frac{\vartheta_i}{\alpha_{2_i}}}}{\gamma_{2_i}^{\frac{\vartheta_i}{\alpha_{2_i}}}}\gamma_{2_i}\bar{\mathcal B}_i(x_i) \!+\! \bar\varphi_{2_i}\Vert w_i\Vert^2 \!+\! \bar\eta_{2_i}\\
			&\overset{\eqref{Eq:13}}{\leq}\!\!\gamma_{2_i}^{\frac{\alpha_{2_i}-\varepsilon_{2_i}}{\alpha_{2_i}}}\mathcal B_i(x_i,\vartheta_i)\!+\! \bar\varphi_{2_i}\Vert w_i\Vert^2 \!+\! \bar\eta_{2_i}. 
		\end{align*}
	  By defining 
          \begin{align*}
              \gamma_i&=\max\{e^{-\gamma_{1_i}\tau}\gamma_{2_i}^{\frac{1}{\alpha_{2_i}}},\gamma_{2_i}^{\frac{\alpha_{2_i}-\varepsilon_{2_i}}{\alpha_{2_i}}}\},\\
              \varphi_i&=\max\{\gamma_{2_i}^{\frac{1}{\alpha_{2_i}}}\!e^{-\gamma_{1_i}\tau} \tau\bar\varphi_{1_i},\bar\varphi_{2_i}\},\\
           \eta_i&=\max\{\gamma_{2_i}^{\frac{1}{\alpha_{2_i}}}\!e^{-\gamma_{1_i}\!\tau}\tau\bar\varphi_{1_i}\Vert w_i\Vert_\infty^2\!+ \gamma_{2_i}^{\frac{1}{\alpha_{2_i}}}\!e^{-\gamma_{1_i}\!\tau}\tau\bar\eta_{1_i},\bar\eta_{2_i}\},
       \end{align*} 
        we have
		\begin{align*}
			\EE \Big[\mathcal B_i(x_i',\!\vartheta_i')\big|x_i,\!\nu_i,\!w_i,\!\vartheta_i\Big]\leq \gamma_i \mathcal B_i(x_i,\!\vartheta_i)+\varphi_{i}\Vert w_i\Vert^2+\eta_i.
		\end{align*} 
To complete the proof, we proceed with showing that $0<\gamma_i<1$ for all Bounds 1--3. One can readily verify that $$0<\gamma_i=\max\{e^{-\gamma_{1_i}\tau},\gamma_{2_i}\}<1,$$ for Bound 1 since $\gamma_{1_i}>0$ and $0<\gamma_{2_i}<1$.
In Bound 2, since  $0<\alpha_{1_i}<1$ and $\gamma_{1_i}>0$, we have $e^{-\gamma_{1_i}\tau(1-\alpha_{1_i})}<1$. In addition, 
\begin{align*}
	e^{-\gamma_{1_i}\tau \alpha_{1_i} \varepsilon_{1_i}}\gamma_{2_i} < 1 \iff \ln(\gamma_{2_i})-\gamma_{1_i}\tau\alpha_{1_i} \varepsilon_{1_i}<0.
\end{align*}
Since $\ln(\gamma_{2_i})-\gamma_{1_i}\tau \vartheta_i<0$ for any $\vartheta_i \in\{\varepsilon_{1_i},\dots,\varepsilon_{2_i}\}$ according to~\eqref{Eq:12}, and by leveraging the continuity property of real numbers, one can
always find some $0<\alpha_{1_i}<1$  (sufficiently close to $1$) such that $\ln(\gamma_{2_i})-\gamma_{1_i}\tau \alpha_{1_i} \varepsilon_{1_i}<0$. Hence, $0<\gamma_i<1$.
Similarly, in Bound~3, one has $\gamma_{2_i}^{\frac{\alpha_{2_i}-\varepsilon_{2_i}}{\alpha_{2_i} }}<1$, since  $\alpha_{2_i}>\varepsilon_{2_i}$ and $0<\gamma_{2_i}<1$. Additionally, 
\begin{align*}
	e^{-\gamma_{1_i}\tau}\gamma_{2_i}^{\frac{1}{\alpha_{2_i}}}<1\iff\ln(\gamma_{2_i})-\gamma_{1_i}\tau \alpha_{2_i} <0.
\end{align*}
Since $\ln(\gamma_{2_i})-\gamma_{1_i}\tau \vartheta_i<0$ for any $\vartheta_i \in\{\varepsilon_{1_i},\dots,\varepsilon_{2_i}\}$ as in~\eqref{Eq:12}, one can conclude $\ln(\gamma_{2_i})-\gamma_{1_i}\tau \alpha_{2_i}<0$ {for some $\alpha_{2_i}>\varepsilon_{2_i}$ (sufficiently close to $\varepsilon_{2_i}$)}, and consequently, $0<\gamma_i<1$. Hence, $\mathcal{B}_i$ as in~\eqref{Eq:13} is an A-CSBC for $\mathcal{A}_i(\Phi_i)$, which completes the proof. $\hfill\blacksquare$

Theorem~\ref{Thm:Main} establishes that the A-CSBC conditions are fulfilled under the transformation $\mathcal{B}_i(x_i,\vartheta_i) = \varpi_i (\vartheta_i)\bar{\mathcal B}_i(x_i)$, provided that condition~\eqref{Eq:12} holds. The proposed approach systematically addresses all hybrid transitions (both flows and jumps) by categorizing them into three distinct cases exhibited by the original CBC.
\begin{remark}
	The use of multiple definitions of control barrier certificates in this work serves distinct yet complementary roles within our framework. Definition~\ref{Def_1a1} introduces A-CSBCs tailored to the A-SHS model defined in Definition~\ref{Def:A-SHS}. While A-CSBCs are necessary to ensure compositional safety using the results derived over subsystems, their design is complicated by the inclusion of the auxiliary counter $\vartheta_i$. In contrast, Definition~\ref{cbc} presents a conventional CBC formulation (excluding the effect of $\vartheta_{i}$) for SHSs by combining the results of stochastic non-hybrid continuous- and discrete-time systems, which serves as the foundational tool for safety analysis. To bridge the gap between these two definitions, Theorem~\ref{Thm:Main} establishes a systematic method to construct A-CSBCs from  CBCs by introducing $\varpi_i(\vartheta_i)$.
\end{remark}

\begin{remark}\label{rmk:novelty}
		The core challenge in analyzing interconnected SHSs lies not only in their high dimensionality but also in the hybrid coupling between subsystems, captured by the disturbance input $w_i$ in~\eqref{Eq:1}. This coupling significantly impacts the system’s behavior and should be explicitly considered in the formal safety analysis. To this end, $w_i$ is integrated in our development in Lemma~\ref{Lemma:1}, its bound is formally established in Theorem~\ref{Thm:2}, and its impact is captured by the parameter $\varphi_i$, introduced in Definition~\ref{Def_1a1}. This parameter is then properly designed in Theorem~\ref{Thm:Main} under three distinct scenarios (Bounds 1--3), that can cover the {hybrid nature} of dynamics.
\end{remark}
\section{Computation of CBC}\label{compute_CBC}
Here, we employ  sum-of-squares (SOS) optimization techniques~\citep{parrilo2003semidefinite} and search for a CBC and its corresponding controller for each $\Phi_i$. {To do so, we assume that $f_{1_i}$ and $f_{2_i}$ are polynomial functions of states $x_i$, disturbances $w_i$, and inputs $\nu_i$. Additionally, $\rho_i$ and $\sigma_i$ are assumed to be polynomial functions of states $x_i$.} The next lemma, inspired by~\cite{nejati2022compositional} (see Lemma~7.1, arXiv version), provides the SOS formulation.

\begin{lemma}\label{sos}
	Suppose sets $X_{0_i},X_{u_i}, X_i, U_i,$  and $W_i$ can be described by $X_{0_i}=\{x_i\in\R^{n_i}\mid g_{0_i}(x_i)\geq0\}$, $X_{u_i}=\{x_i\in\R^{n_i}\mid g_{u_i}(x_i)\geq0\}$, $X_i=\{x_i\in\R^{n_i}\mid g_i(x_i)\geq0\}$, {$U_i=\{\nu_i\in\R^{m_i}\mid g_{\nu_i}(\nu_i)\geq0\}$,} and {$W_i=\{w_i\in\R^{p_i}\mid g_{w_i}(w_i)\geq0\}$}.
	Let there exist an SOS polynomial $\bar{\mathcal B}_i(x_i)$, constants $\gamma_{1_i}\in\R,\bar\kappa_i,\gamma_{2_i}\in\R^+$, $\bar\mu_i,\bar\beta_i,\bar\varphi_{1_i},\bar\varphi_{2_i},\bar\eta_{1_i}, \bar\eta_{2_i}\in\R_{0}^+$, and polynomial $l_{\nu_{j_i}}(x_i)$, corresponding to the $j^{\text{th}}$ input in $\nu_i=[\nu_{1_i};\ldots;\nu_{m_i}]\in U_i\subseteq \R^{m_i}$. Suppose there exist vectors of SOS polynomials $l_{0_i}(x_i)$, $l_{u_i}(x_i)$, $\tilde l_{i}(x_i)$, $l_i(x_i,\nu_i,w_i)$, $\hat l_i(x_i,\nu_i,w_i)$, $l_{\nu_i}(x_i,\nu_i,w_i)$, $l_{w_i}(x_i,\nu_i,w_i)$, and $\hat l_{w_i}(x_i,\nu_i,w_i)$ of appropriate dimensions such that the following expressions are SOS polynomials:
	\begin{subequations}
		\begin{align}\label{eq:sos11}
			&\hphantom{-}~~\bar{\mathcal B_i}(x_i)-\tilde l_{i}^\top(x_i) g_{i}(x_i)- \bar\kappa_ix_i^\top x_i,\\\label{eq:sos1}
			&-\bar{\mathcal B_i}(x_i)-l_{0_i}^\top(x_i) g_{0_i}(x_i)+\bar\mu_i,\\\label{eq:sos2}
			&\hphantom{-}~~\bar{\mathcal B}_i(x_i)\!-\!l_{u_i}^\top(x_i) g_{u_i}(x_i)\!-\!\bar\beta_i ,\\\notag
			&-\mathcal{L}\bar{\mathcal B}_i(x_i) \!-\! \gamma_{1_i} \!\bar{\mathcal B}_i(x_i)\! \!+\!\bar\varphi_{1_i}w_i^\top w_i\!+\!\bar\eta_{1_i}\!
			\!-\!\!\sum_{j=1}^{m_i}(\nu_{j_i}\!\!-\!l_{\nu_{j_i}}\!\!(x_i)\!)\\\notag
			&- l_i^\top(x_i,\nu_i,w_i) g_i(x_i)-l_{\nu_i}^\top(x_i,\nu_i,w_i) g_{\nu_i}(\nu_i)\\\label{eq:sos3}
			& - l_{w_i}^\top(x_i,\nu_i,w_i) g_{w_i}(w_i),\\\notag
			&-\EE\Big[\bar{\mathcal B}_i(f_{2_i}\!(x_i,\nu_i,w_i,{\varsigma_i})\!) \big| x_i,\!\nu_i,\!w_i\Big]\!\!+\! \gamma_{2_i}\bar{\mathcal B}_i(x_i) \!+\! \bar\varphi_{2_i}w_i^\top \!w_i \\\notag
			&+ \bar\eta_{2_i}\!-\!\sum_{j=1}^{m_i}(\nu_{j_i}\!-\! l_{\nu_{j_i}}\!\!(x_i)\!)- \hat l_i^\top(x_i,\nu_i,w_i) g_i(x_i)\\\label{eq:sos4}
			&- l_{\nu_i}^\top(x_i,\nu_i,w_i) g_{\nu_i}(\nu_i)- \hat l_{w_i}^\top(x_i,\nu_i,w_i) g_{w_i}(w_i).
		\end{align}
	\end{subequations}
	Then $\bar{\mathcal B}_i(x_i)$ fulfills conditions~\eqref{subsys21}--\eqref{csbceq} in Definition~\ref{cbc}. Furthermore, any $\nu_{j_i} \geq l_{\nu_{j_i}}(x_i)$ is the corresponding safety controller.
\end{lemma}
	
\section{Case Study}\label{Sec:Case}
Here, we demonstrate the effectiveness of our results by applying them to an interconnected stochastic hybrid system, composed of $1000$ nonlinear subsystems under two distinct interconnection topologies, namely, a ring topology and a $3$-in-$3$-out topology.
\subsection{Ring Interconnection Topology}\label{subsec:ring}
In this subsection, we apply our results to an interconnected SHS $\Phi$ composed of $1000$ nonlinear subsystems described by
\begin{align}\label{Eq:14}
	\Phi\!:\left\{\hspace{-1.5mm}
	\begin{array}{rl}
		\mathsf{d}x(t)\!=\!\!& (A_1x^{\circ 3}(t)\!+\!B_1\nu(t))\mathsf{d}t + E\mathsf{d}\mathbb W_t+F\mathsf{d}\mathbb P_t,\\
		&~\quad\quad\quad\quad\quad\quad\quad\quad\quad\quad\quad\quad \quad ~ t\in\mathbb{R}_{0}^+\backslash \Delta,\\
		x(t)\!=\!\!& A_2x^{\circ 3}(t^-)\!+\! B_2\nu(t)\!+\!G\varsigma(t), \quad ~~~\!t\in \Delta,
	\end{array}
	\right.
\end{align}
where  $x=[x_1;\ldots;x_{1000}]$, and $\nu=[\nu_1;\ldots;\nu_{1000}]$. The matrices $A_1$ and $A_2$ have diagonal entries given by $\bar a_{1_{ii}}=a_1$ and $\bar a_{2_{ii}}=a_2$, respectively. Their only nonzero off-diagonal entries are $\bar a_{1_{i,i-1}}=\bar a_{1_{1,1000}}=d_1$ and $\bar a_{2_{i,i-1}}=\bar a_{2_{1,1000}}=d_2$ for all $i\in \{2,\ldots,1000\}$, while all other entries are zero. Moreover, $B_1=b_1\mathds I_{1000}$, $B_2=b_2\mathds I_{1000}$, $E = 0.6\mathds{I}_{1000}, F = 0.5\mathds{I}_{1000},$ and $G = 0.5\mathds{I}_{1000}$. 

Now, by considering individual subsystems $\Phi_i$ as
\begin{align}\label{Eq:141}
	\Phi_i\!\!:\!\left\{\hspace{-1.5mm}
	\begin{array}{rl}
		\mathsf{d}x_i(t)\!=\!\!& (a_1x_i^3(t)\!+\!b_1\nu_i(t)\!+\!d_1w_i(t))\mathsf{d}t+ 0.6\mathsf{d}\mathbb W_t\\
		\quad\quad &+ ~0.5\mathsf{d}\mathbb P_t,\quad\quad\quad\quad\quad\quad\quad\quad\quad t\in\mathbb{R}_{0}^+\backslash \Delta_i,\\
		x_i(t)\!=\!& a_2x_i^3(t^-)\!+\! b_2\nu_i(t)\!+\!d_2w_i(t^-)\!+\!0.5\varsigma_i(t),\\
		& \quad\quad\quad\quad\quad\quad\quad\quad\quad\quad\quad\quad\quad~ t\!\in\! \Delta_i,
	\end{array}
	\right.
\end{align}
with $a_1,b_1,a_2,b_2 \in \mathbb R$, jump parameters $\tau= 0.1$, $\varepsilon_{1_i} = 1,\varepsilon_{2_i} = 7$, and rates of the Poisson process as $\lambda_i = 0.5$, one can readily see that $\Phi =\mathcal{I}(\Phi_1,\ldots,\Phi_{1000})$, where $w_i(t)=x_{i-1}(t)$ with $x_0=x_{1000}$, has a ring interconnection topology.

The regions of interest are given as $X = [0,10]^{1000}, X_{0} = [0,2]^{1000},$ and $X_{u} = [9,10]^{1000}$. The main goal is to design an A-CBC for the interconnected A-SHS and its corresponding safety controller based on A-CSBCs of individual subsystems such that the {states of $\Phi$ in~\eqref{Eq:14} do not reach the unsafe set $X_{u} = [9,10]^{1000}$}. To do so, we first search for CBCs and accordingly design local controllers for $\Phi_i$. Subsequently, A-CSBCs are constructed from CBCs according to Theorem~\ref{Thm:Main}. We illustrate the obtained results for different values of $a_1$, $b_1$, $a_2$, and $b_2$. We employ \textsf{SOSTOOLS}~\citep{papachristodoulou2013sostools} and the SDP solver \textsf{SeDuMi}~\citep{sturm1999using} to compute CBCs.

For $a_1 = -0.4,b_1 = 0.5,a_2 = 0.01,$ $b_2 = 0.6$, we compute a CBC of order four for each subsystem as $\bar{\mathcal B_i}(x_i) = 0.0027x_i^4 - 0.017x_i^3 + 0.0408x_i^2 - 0.0442 x_i + 0.03241$ and its corresponding controller as $\nu_i = -0.09x_i + 2.7$, based on Lemma~\ref{sos}. Moreover, the corresponding constants in Definition~\ref{cbc} satisfying conditions~\eqref{subsys21}--\eqref{csbceq} are quantified as $\bar\mu_i = 0.0675$, $\bar\beta_i = 7.2187$, $\bar\kappa_i = 0.0011$, $\bar\eta_{1_i} = 0.0015$, $\bar\eta_{2_i} = 0.0012$, $\bar\varphi_{1_i} = 0.00001$, $\bar\varphi_{2_i} = 0.00001$, $\gamma_{1_i} = 0.1$, and $\gamma_{2_i} = 0.99$.
We now select $\alpha_{1_i} = 0.1$ and $\alpha_{2_i} = 8$, and employ Theorem~\ref{Thm:Main} to construct an A-CSBC for the augmented SHS $\mathcal A_i(\Phi_i)$ using the obtained CBC. Since  $\gamma_{1_i} =  0.1>0$ and $0<\gamma_{2_i} = 0.99 <1$, the results of Bound 1 are useful. One can readily verify that condition~\eqref{Eq:12} is also met for all $\vartheta_i \in \{1,\ldots,7\}$. Then, one can conclude that $\mathcal B_i(x_i,\vartheta_i)\coloneq\bar{\mathcal B}_i(x_i)$ is an A-CSBC for $\mathcal A_i(\Phi_i)$ with $\gamma_i=\max\{e^{-\gamma_{1_i}\tau},\gamma_{2_i}\} =\max\{0.99, 0.99 \}=0.99$, $\varphi_i=\max\{\tfrac{(1-e^{-\gamma_{1_i}\!\tau})}{\gamma_{1_i}} \bar\varphi_{1_i},\bar\varphi_{2_i}\}=\max\{9.9052 \times 10^{-7}, 10^{-5}\}=10^{-5}$\!, and $\eta_i =\max\{\tfrac{(1-e^{-\gamma_{1_i}\!\tau})}{\gamma_{1_i}} \bar\varphi_{1_i} \Vert w_i\Vert_\infty^2 + \tfrac{(1-e^{-\gamma_{1_i}\!\tau})}{\gamma_{1_i}} \bar\eta_{1_i}, \\ \bar\eta_{2_i}\} =\max\{2.2985 \times 10^{-4}, 0.0012\}= 0.0012$. We now proceed to construct an A-CBC based on A-CSBCs of A-SHSs according to Theorem~\ref{Thm:2}. By selecting $\xi_i= 0.001$, $\forall i\in \{1,\dots,1000\}$, one has~$\xi_i\Gamma_i = -1.056 \times \! 10^{-6} \!<\! 0,$ implying the satisfaction of condition~\eqref{Eq:21}. Moreover, we compute $\beta =\sum_{i=1}^{1000}\xi_i \beta_i = 7.2187$, and $\mu = \sum_{i=1}^{1000}\xi_i \mu_i= 0.0675$, which yield condition~\eqref{Eq:28} being satisfied. Then, it can be concluded that $\mathcal B(x,\vartheta) = \sum_{i=1}^{1000}\xi_i\mathcal B_i(x_i,\vartheta_i),$ satisfying conditions~\eqref{Eq_2a1}--\eqref{Eq_3a} with $\eta = \sum_{i=1}^{1000}\xi_i\eta_i=0.0012,$ and $\gamma = 1 + \Gamma = 0.9989$.

\begin{figure*}
	\centering
	\includegraphics[width=0.3\linewidth]{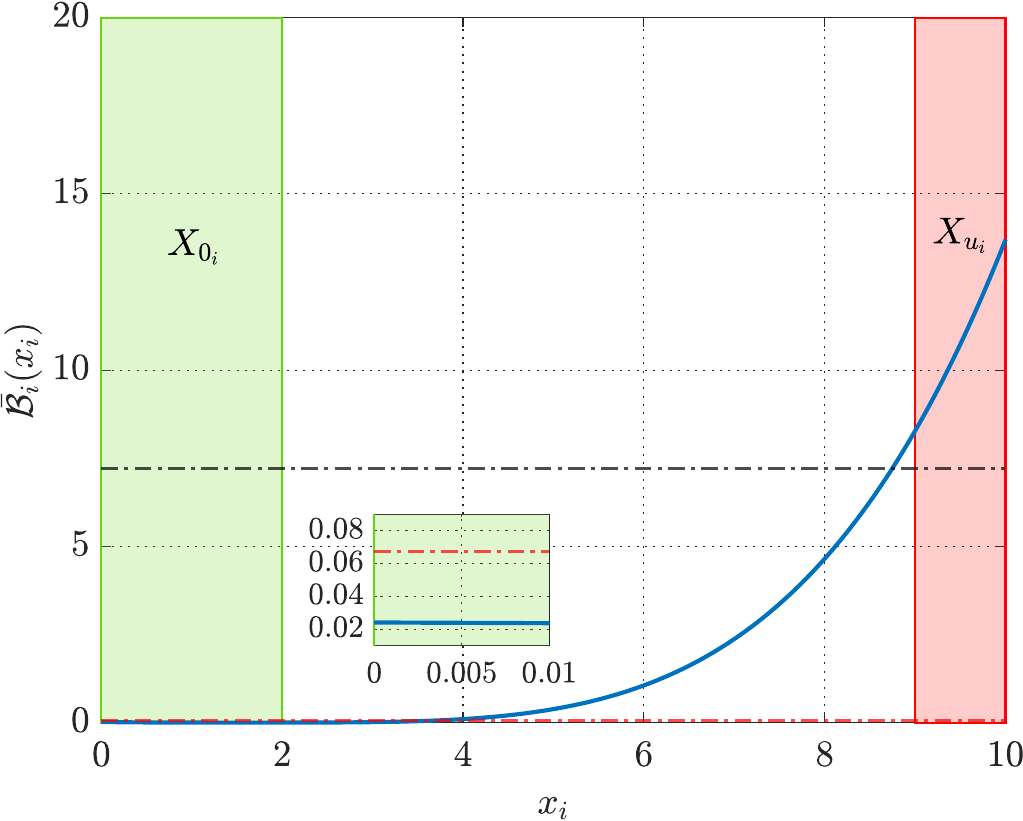}\hspace{0.5cm}
	\includegraphics[width=0.3\linewidth]{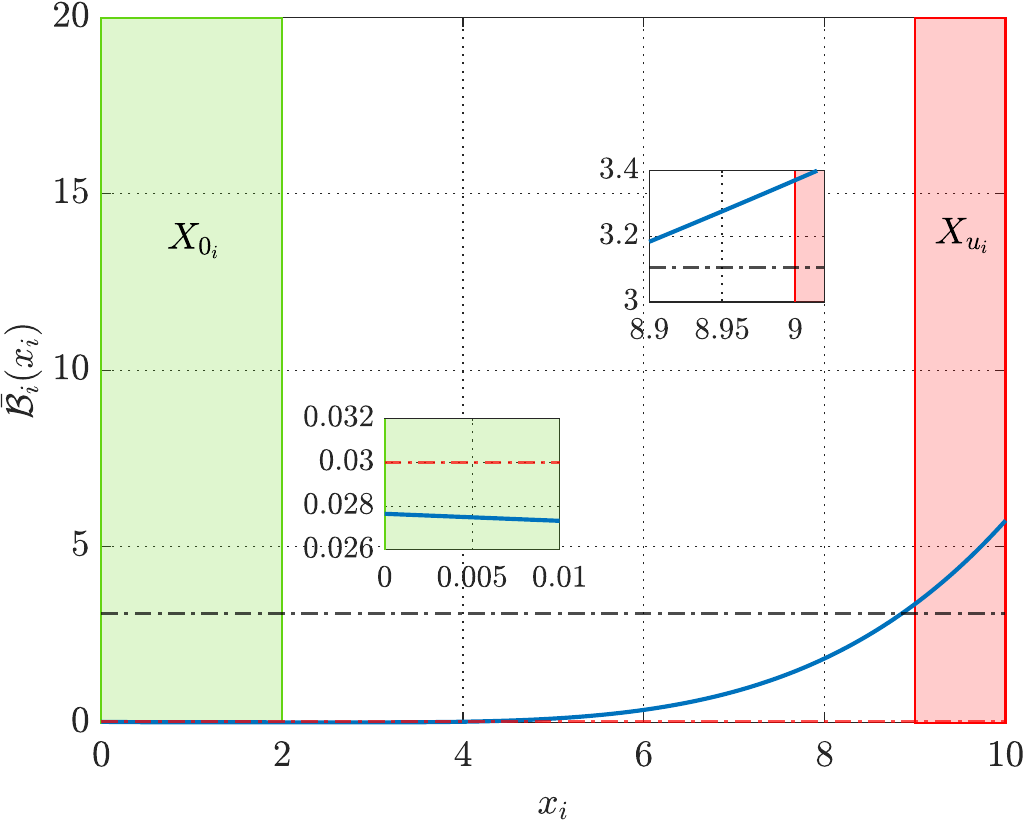}\hspace{0.5cm}
	\includegraphics[width=0.3\linewidth]{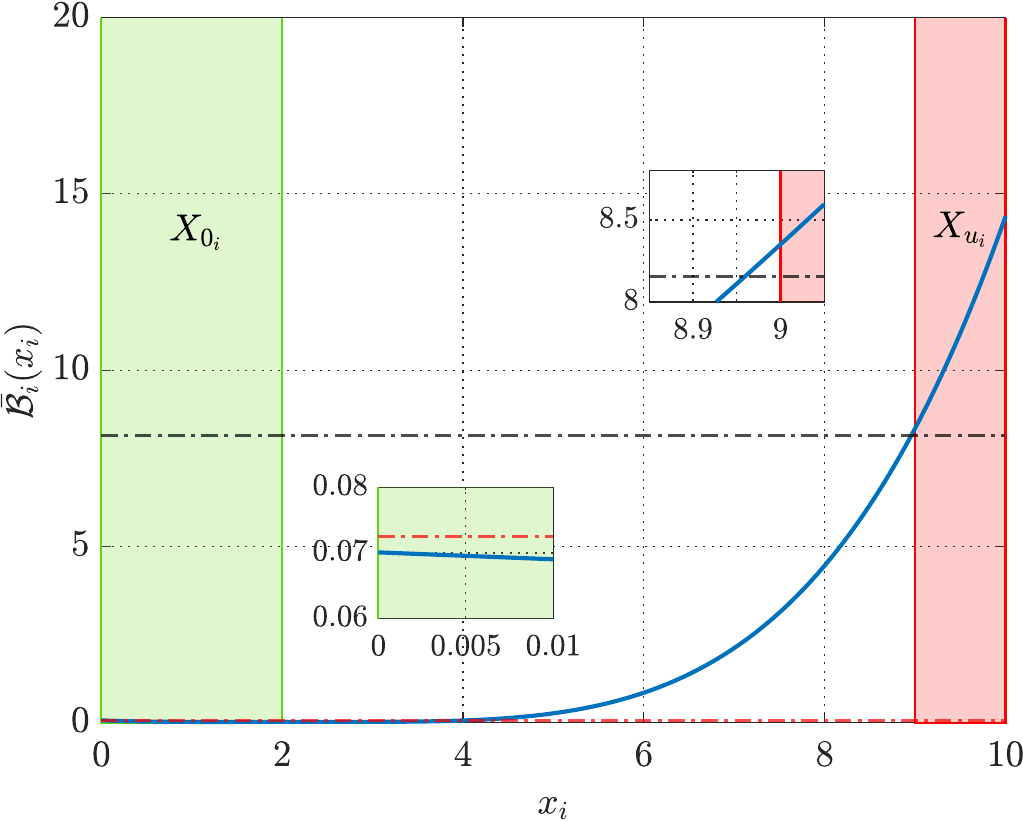}\\
	\includegraphics[width=0.3\linewidth]{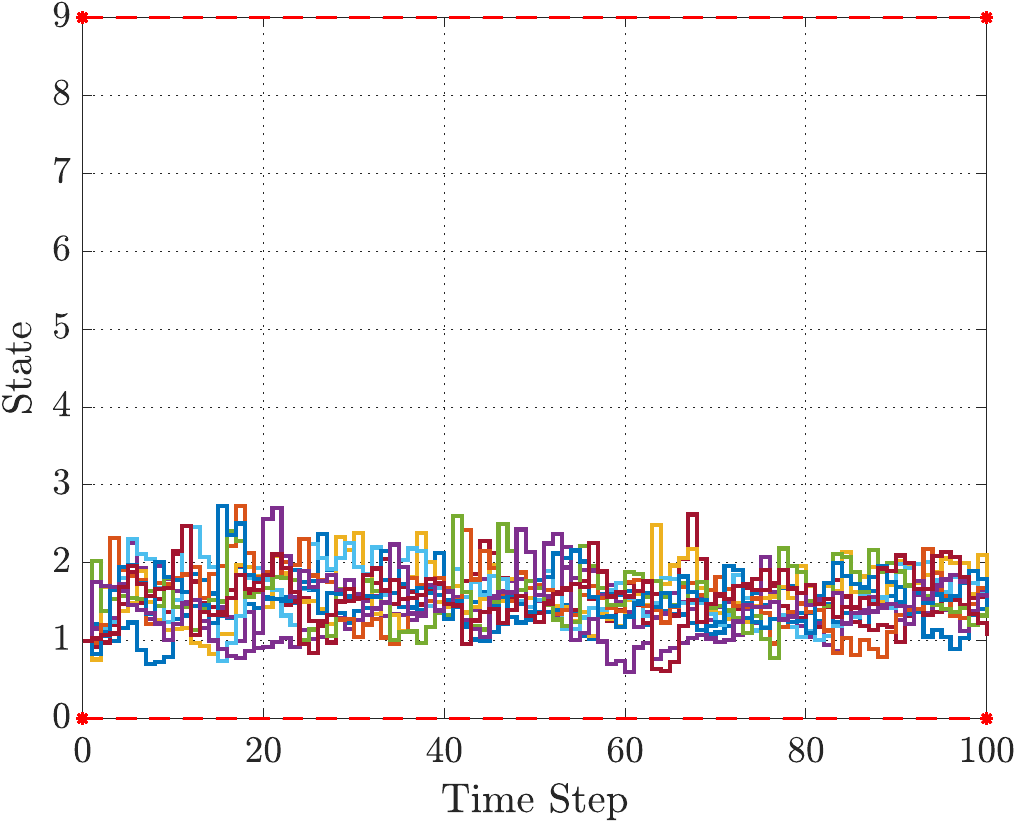}\hspace{0.5cm}
	\includegraphics[width=0.3\linewidth]{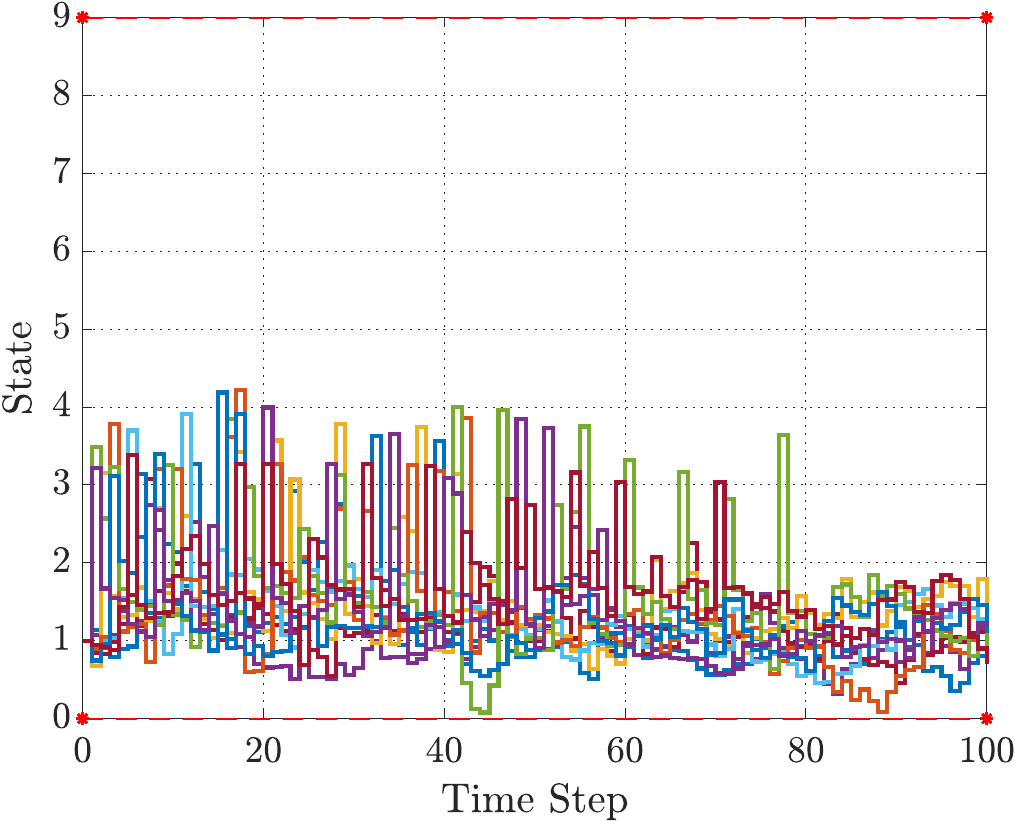}\hspace{0.5cm}
	\includegraphics[width=0.3\linewidth]{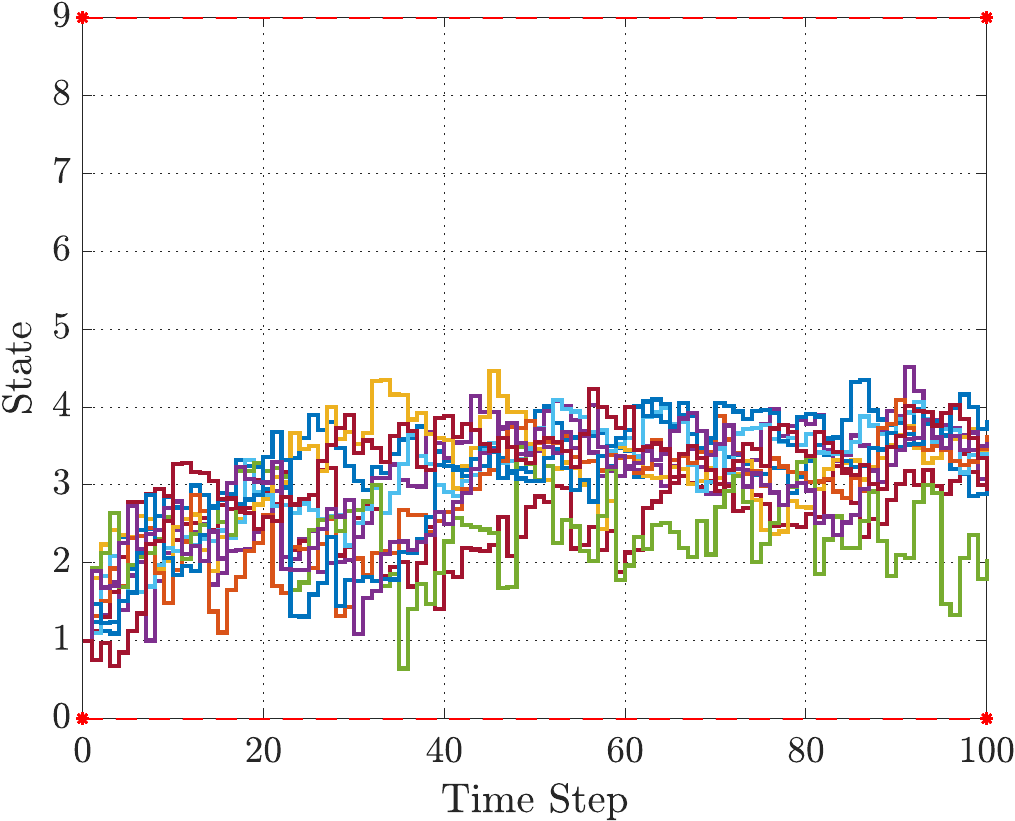}
	\caption{Control barrier certificate \sampleline{mycolor, thick} (top) and closed-loop state trajectories (bottom) of a representative subsystem of the interconnected SHS $\Phi$ with ring topology, under 10 noise realizations, for  $a_1 = -0.4,b_1 = 0.5,a_2 = 0.01,b_2 = 0.6$ (left), $a_1 = -0.3,b_1 = 0.2,a_2 = 1.01,b_2 = 1$ (middle), and $a_1 = 0.01,b_1 = 0.7,a_2 = 0.075,b_2 = 0.8$ (right). Initial and unsafe regions are depicted by green \protect\greensquare\ and red \protect\redsquare\ boxes. Moreover, $\bar {\mathcal B}_i(x_i) = \bar {\mu}_i$ and $\bar {\mathcal B}_i(x_i) = \bar {\beta}_i$ are indicated by~\sampleline{dash dot, red, thick} and~\sampleline{dash dot, black , thick}, respectively. {Here, jump instants are assumed to be identical among all subsystems.}}
	\label{fig:w1}
\end{figure*}

For $a_1 = -0.3,b_1 = 0.2,a_2 = 1.01,b_2 = 1$, we compute a CBC of order four as $\bar{\mathcal B}_i(x_i) = 0.0014  x^4_i - 0.0103  x^3_i + 0.0277  x^2_i - 0.0324 x_i + 0.0277$ and its corresponding controller as $\nu_i = -1.02x^3_i + 3.05$. Furthermore, the associated constants satisfying conditions~\eqref{subsys21}--\eqref{csbceq} are obtained as $\bar\mu_i = 0.03$, $\bar\beta_i = 3.1046$, $\bar\kappa_i =  9.4773 \times 10^{-4}$, $\gamma_{1_i} = 0.451$, $\gamma_{2_i} = 1.00001$, $\bar\varphi_{1_i} = 1.3384 \times 10^{-5}$, $\bar\varphi_{2_i} = 2.5902 \times 10^{-6}$, $\bar\eta_{1_i} =  0.031$, and $\bar\eta_{2_i} = 0.003$. Since  $\gamma_{1_i} =  0.451>0$ and $\gamma_{2_i} = 1.00001 \geq1$, Bound 2 is valid. As condition~\eqref{Eq:12} is met for all $\vartheta_i \in \{1,\ldots,7\}$, we obtain an A-CSBC for $\mathcal A_i(\Phi_i)$ as $\mathcal B_i(x_i,\vartheta_i)\coloneq e^{4.5 \times 10^{-3} \vartheta_i}\bar{\mathcal B}_i(x_i)$ with $\mu_i = 0.03$, $\beta_i = 3.1046$, $\gamma_i=\max\{e^{-\gamma_{1_i}\tau(1-\alpha_{1_i})},e^{-\gamma_{1_i}\tau \alpha_{1_i} \varepsilon_{1_i}}\gamma_{2_i}\} =\max\{  0.9602, 0.9955\} =0.9955$, $\varphi_i=\max\{e^{\gamma_{1_i}\tau \alpha_{1_i} \varepsilon_{2_i}}$ $\tfrac{(1-e^{-\gamma_{1_i}\!\tau})}{\gamma_{1_i}} \bar\varphi_{1_i},\bar\varphi_{2_i}\} =\max\{1.3507\times 10^{-6}, 2.5902 \times 10^{-6}\} = 2.5902 \times 10^{-6}$, and $\eta_i\!=\! \max\{e^{\gamma_{1_i}\tau \alpha_{1_i}\varepsilon_{2_i}}$ $\tfrac{(1-e^{-\gamma_{1_i}\tau})}{\gamma_{1_i}}\bar{\varphi}_{1_i}\,\|w_i\|_\infty^2 \!+\!e^{\gamma_{1_i}\tau \alpha_{1_i}\varepsilon_{2_i}}\tfrac{(1-e^{-\gamma_{1_i}\tau})}{\gamma_{1_i}}\bar{\eta}_{1_i},\,\bar{\eta}_{2_i}\}=\max\{0.0032, 0.003\}=0.0032.$ We now proceed to construct an A-CBC for $\mathcal A(\Phi)$ based on the obtained A-CSBCs according to Theorem~\ref{Thm:2}. By choosing $\xi_i= 0.001,$~$\forall i\in \{1,\dots,1000\}$, one has~$\xi_i\Gamma_i = -1.7568 \times \! 10^{-6} \!<\! 0,$ implying the satisfaction of condition~\eqref{Eq:21}. Moreover, we compute $\beta =\sum_{i=1}^{1000}\xi_i \beta_i = 3.2042$, and $\mu =\sum_{i=1}^{1000}\xi_i \mu_i= 0.0301$, implying the fulfillment of condition~\eqref{Eq:28}. Hence, it can be concluded that $\mathcal B(x,\vartheta) = \sum_{i=1}^{1000}\xi_i\mathcal B_i(x_i,\vartheta_i)$ satisfies conditions~\eqref{Eq_2a1}--\eqref{Eq_3a} with $\eta = \sum_{i=1}^{1000}\xi_i\eta_i= 0.0032$ and $\gamma =  1 + \Gamma = 0.9982.$

For $a_1 = 0.01,b_1 = 0.7,a_2 = 0.075,b_2 = 0.8$, we compute a CBC of order four as $\bar{\mathcal B}_i(x_i) =  0.0037x^4_i - 0.02992x^3_i + 0.0875x^2_i - 0.1062x_i + 0.07$ and its corresponding controller as $\nu_i = -0.099x^3_i + 2.5$. In addition, the associated constants in Definition~\ref{cbc} are acquired as $\bar\mu_i = 0.0725$, $\bar\beta_i = 8.1535$, $\bar\kappa_i = 0.0022$, $\gamma_{1_i} = -0.21$, $\gamma_{2_i} = 0.77$, $\bar\varphi_{1_i} = 7.3679 \times 10^{-6}$, $\bar\varphi_{2_i} = 7.3395 \times 10^{-6}$, and $\bar\eta_{1_i} = 0.01$, and $\bar\eta_{2_i} = 0.01$. Since  $\gamma_{1_i} = -0.21\leq0$ and $0<\gamma_{2_i} = 0.77 <1$, the results of Bound 3 can be employed. Since condition~\eqref{Eq:12} is also met for all $\vartheta_i \in \{1,\ldots,7\}$, one can conclude that $\mathcal B_i(x_i,\vartheta_i)\coloneq 0.77^{\frac{\vartheta_i}{8}}\bar{\mathcal B_i}(x_i)$ is an A-CSBC for $\mathcal A_i(\Phi_i)$ with
$\mu_i = 0.0725$, $\beta_i = 6.4867$, $\gamma_i=\max\{e^{-\gamma_{1_i}\tau}\gamma_{2_i}^{\frac{1}{\alpha_{2_i}}},\gamma_{2_i}^{\frac{\alpha_{2_i}-\varepsilon_{2_i}}{\alpha_{2_i}}}\} =\max\{0.9884,     0.9679\}=0.9884$, $\varphi_i=\max\{\gamma_{2_i}^{\frac{1}{\alpha_{2_i}}}e^{-\gamma_{1_i}\tau} \tau\bar\varphi_{1_i},\bar\varphi_{2_i}\} =\max\{ 7.2824 \times 10^{-7},  7.3395 \times 10^{-6}\}=7.3395 \times 10^{-6}$, and $\eta_i=\max\{\gamma_{2_i}^{\frac{1}{\alpha_{2_i}}}e^{-\gamma_{1_i}\tau} \tau\bar\varphi_{1_i}\Vert w_i\Vert_\infty^2 + \gamma_{2_i}^{\frac{1}{\alpha_{2_i}}}e^{-\gamma_{1_i}\tau}\tau\bar\eta_{1_i},\bar\eta_{2_i}\} =\max\{ 0.001, 0.01\}= 0.01$. We now proceed to construct an A-CBC for $\mathcal A(\Phi)$ based on the synthesized A-CSBCs according to Theorem~\ref{Thm:2}. By selecting $\xi_i= 0.001$, $\forall i\in \{1,\dots,1000\}$, one has $\xi_i\Gamma_i = -7.3813 \times \! 10^{-6} \!<\! 0$, implying the satisfaction of condition~\eqref{Eq:21}. Moreover, we compute $\beta =\sum_{i=1}^{1000}\xi_i \beta_i =  7.8914$, and $\mu = \sum_{i=1}^{1000}\xi_i \mu_i= 0.0577$, resulting in the fulfillment of condition~\eqref{Eq:28}. Then, it can be concluded that $\mathcal B(x,\vartheta) = \sum_{i=1}^{1000}\xi_i\mathcal B_i(x_i,\vartheta_i)$, satisfying conditions~\eqref{Eq_2a1}--\eqref{Eq_3a} with $\eta = \sum_{i=1}^{1000}\xi_i\eta_i=0.01,$ and $\gamma =1 + \Gamma = 0.9926.$

By employing Theorem~\ref{Kushner}, we guarantee that the state trajectory of $\Phi$ starting from initial conditions inside $X_{0} = [0,2]^{1000}$ {avoids the unsafe set $X_u=[9,10]^{1000}$} during $\mathcal T=100$ time steps with probabilities of at least $0.9743$ (Bound~1), $0.8922$ (Bound~2), and $0.8474$ (Bound~3)\footnote{The reported safety probabilities were computed using full-precision values.}. Fig.~\ref{fig:w1} depicts the synthesized CBC for a representative subsystem $\Phi_i$ together with the corresponding closed-loop state trajectories under Bounds~1--3 and 10 noise realizations. Consistent with the guaranteed safety intervals, the trajectories are shown only at discrete time instants. The computation of the CBC and its corresponding controller for each subsystem (regarding each bound) took almost $4.81$ seconds with a memory usage of $64.84~\mathsf{Mbit}$ on a MacBook Pro M2 Max chip with $32$ GB of memory.

\subsection{3-in-3-out Interconnection Topology}\label{subsec:uniform}
In this subsection, we consider an interconnected network comprising 1000 nonlinear subsystems, where each subsystem is influenced by three preceding subsystems and affects three following ones. The model is the same as in \eqref{Eq:14}, where the diagonal elements of the matrices $A_1$ and $A_2$ are given by $\bar{a}_{1_{ii}} = a_1$ and $\bar{a}_{2_{ii}} = a_2$, respectively. Their off-diagonal elements are defined as
$
\bar{a}_{1_{i,i-1}} = \bar{a}_{1_{i,i-2}} = \bar{a}_{1_{i,i-3}} = \bar{a}_{1_{1,1000}} = \bar{a}_{1_{1,999}} = \bar{a}_{1_{1,998}} = \bar{a}_{1_{2,1}} = \bar{a}_{1_{2,1000}} = \bar{a}_{1_{2,999}} = \bar{a}_{1_{3,1}} = \bar{a}_{1_{3,2}} = \bar{a}_{1_{3,1000}} = d_1,
$
$
\bar{a}_{2_{i,i-1}} = \bar{a}_{2_{i,i-2}} = \bar{a}_{2_{i,i-3}} = \bar{a}_{2_{1,1000}} = \bar{a}_{2_{1,999}} = \bar{a}_{2_{1,998}} = \bar{a}_{2_{2,1}} = \bar{a}_{2_{2,1000}} = \bar{a}_{2_{2,999}} = \bar{a}_{2_{3,1}} = \bar{a}_{2_{3,2}} = \bar{a}_{2_{3,1000}} = d_2,
$
for all $i \in \{4, \ldots, 1000\}$.
Each subsystem in this network is described by
\begin{align}\label{Eq:141_1}
	\Phi_i\!\!:\!\left\{\hspace{-1.5mm}
	\begin{array}{rl}
		\mathsf{d}x_i(t)\!=\!\!& (a_1x_i^3(t)\!+\!b_1\nu_i(t)\!+\!D_1w_i(t))\mathsf{d}t+ 0.6\mathsf{d}\mathbb W_t\\
		\quad\quad &+ ~0.5\mathsf{d}\mathbb P_t,\quad\quad\quad\quad\quad\quad\quad\quad\quad t\in\mathbb{R}_{0}^+\backslash \Delta_i,\\
		x_i(t)\!=\!& a_2x_i^3(t^-)\!+\! b_2\nu_i(t)\!+\!D_2w_i(t^-)\!+\!0.5\varsigma_i(t),\\
		& \quad\quad\quad\quad\quad\quad\quad\quad\quad\quad\quad\quad\quad~ t\!\in\! \Delta_i,
	\end{array}
	\right.
\end{align}
where $D_1 = [d_1; d_1; d_1]^\top$, $D_2 = [d_2; d_2; d_2]^\top$, and $w_i = [x_{i - 1}; x_{i - 2}; x_{i - 3}]$, with the wrap-around conditions $x_0 = x_{1000}$, $x_{-1} = x_{999}$, and $x_{-2} = x_{998}$.  We refer to this configuration as the \emph{3-in-3-out interconnection topology}, in which the size of $w_i$ is three times that in the \emph{ring} topology. Hence, this represents a denser interconnection structure compared to the one presented in Section~\ref{subsec:ring}. For the purpose of comparing the two topologies, all system parameters are kept identical to those used in Section~\ref{subsec:ring}.

For $a_1 = -0.4,b_1 = 0.5,a_2 = 0.01,$ $b_2 = 0.6$, we synthesize a CBC and its associated safety controller as $\bar{\mathcal B_i}(x_i) = 0.0029x_i^4 - 0.0163x_i^3 + 0.0327x_i^2 - 0.0282 x_i + 0.074$ and $\nu_i = -0.08x_i + 2.6$, respectively. The parameters satisfying conditions~\eqref{subsys21}--\eqref{csbceq} in Definition~\ref{cbc} are designed as $\bar\mu_i = 0.0755$, $\bar\beta_i = 7.6247$, $\bar\kappa_i = 0.009$, $\bar\eta_{1_i} = 0.015$, $\bar\eta_{2_i} = 0.0012$, $\bar\varphi_{1_i} = 0.00001$, $\bar\varphi_{2_i} = 0.00001$, $\gamma_{1_i} = 0.1$, and $\gamma_{2_i} = 0.99$.
By selecting $\alpha_{1_i}=0.1$ and $\alpha_{2_i}=8$, and since $\gamma_{1_i}=0.1>0$ and $0<\gamma_{2_i}=0.99<1$, Bound~1 in Theorem~\ref{Thm:Main} applies. Since condition~\eqref{Eq:12} is fulfilled for all $\vartheta_i\in\{1,\ldots, 7\}$, we obtain $\mathcal B_i(x_i,\vartheta_i)\coloneq\bar{\mathcal B}_i(x_i)$ as an A-CSBC for each A-SHS $\mathcal A_i(\Phi_i)$, with $\mu_i=0.0755$, $\beta_i=7.6247$, $\gamma_i=\max\{e^{-\gamma_{1_i}\tau},\gamma_{2_i}\} =\max\{  0.99,    0.99\}= 0.99$, $\varphi_i=\max\{\tfrac{(1-e^{-\gamma_{1_i}\!\tau})}{\gamma_{1_i}} \bar\varphi_{1_i},\bar\varphi_{2_i}\}= \max\{ 9.9502 \times 10^{-7}, 10^{-5}\}=10^{-5}$ and $\eta_i =\max\{\tfrac{(1-e^{-\gamma_{1_i}\!\tau})}{\gamma_{1_i}} \bar\varphi_{1_i} \Vert w_i\Vert_\infty^2 + \tfrac{(1-e^{-\gamma_{1_i}\!\tau})}{\gamma_{1_i}} \bar\eta_{1_i},\\ \bar\eta_{2_i}\} =\max\{ 0.0017,0.0012\} = 0.0017$. As per Theorem~\ref{Thm:2}, $\xi_i\Gamma_i = -6.63 \times \! 10^{-6} \!<\! 0$ satisfies condition~\eqref{Eq:21} with $\xi_i= 0.001$,~$\forall i\in \{1,\dots,1000\}$. Additionally, $\beta =\sum_{i=1}^{1000}\xi_i \beta_i = 7.6247$ and $\mu = \sum_{i=1}^{1000}\xi_i \mu_i= 0.0755$ fulfill condition~\eqref{Eq:28}. Therefore, $\mathcal B(x,\vartheta) = \sum_{i=1}^{1000}\xi_i\mathcal B_i(x_i,\vartheta_i)$ is an A-CBC for $\mathcal A(\Phi)$, satisfying conditions~\eqref{Eq_2a1}--\eqref{Eq_3a}, with $\eta = \sum_{i=1}^{1000}\xi_i\eta_i=0.0017$ and $\gamma = 1 + \Gamma = 0.9934$.

\begin{figure*}
	\centering
	\includegraphics[width=0.3\linewidth]{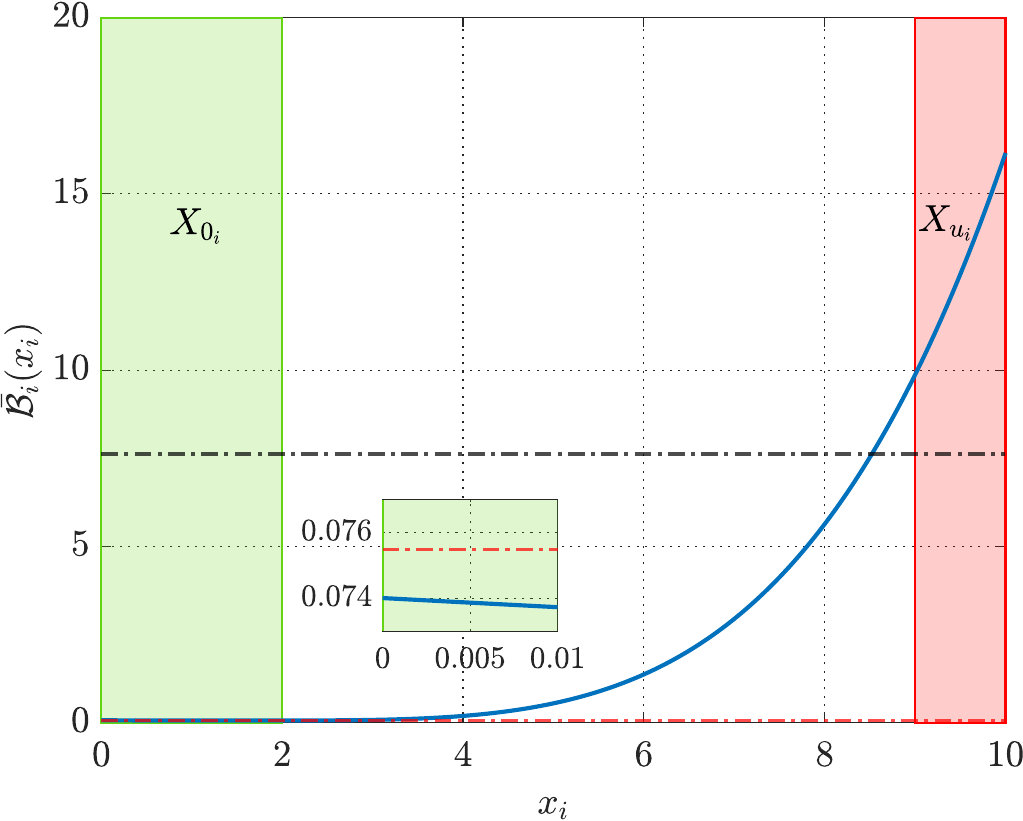}\hspace{0.5cm}
	\includegraphics[width=0.3\linewidth]{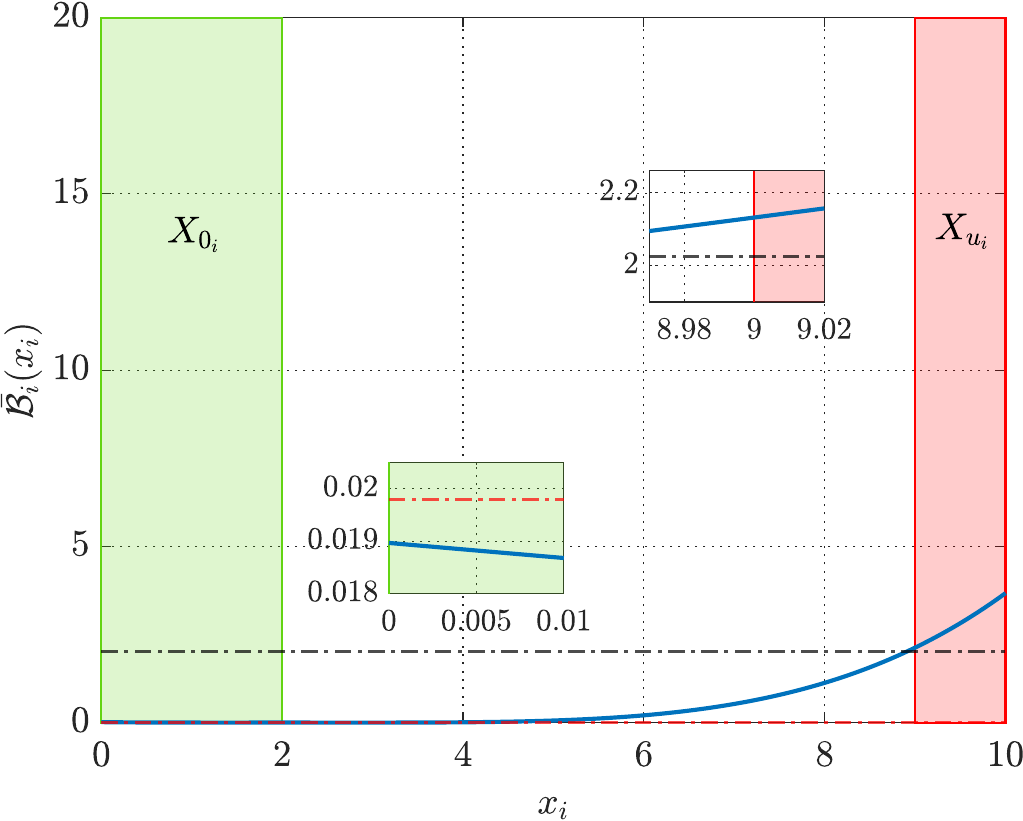}\hspace{0.5cm}
	\includegraphics[width=0.3\linewidth]{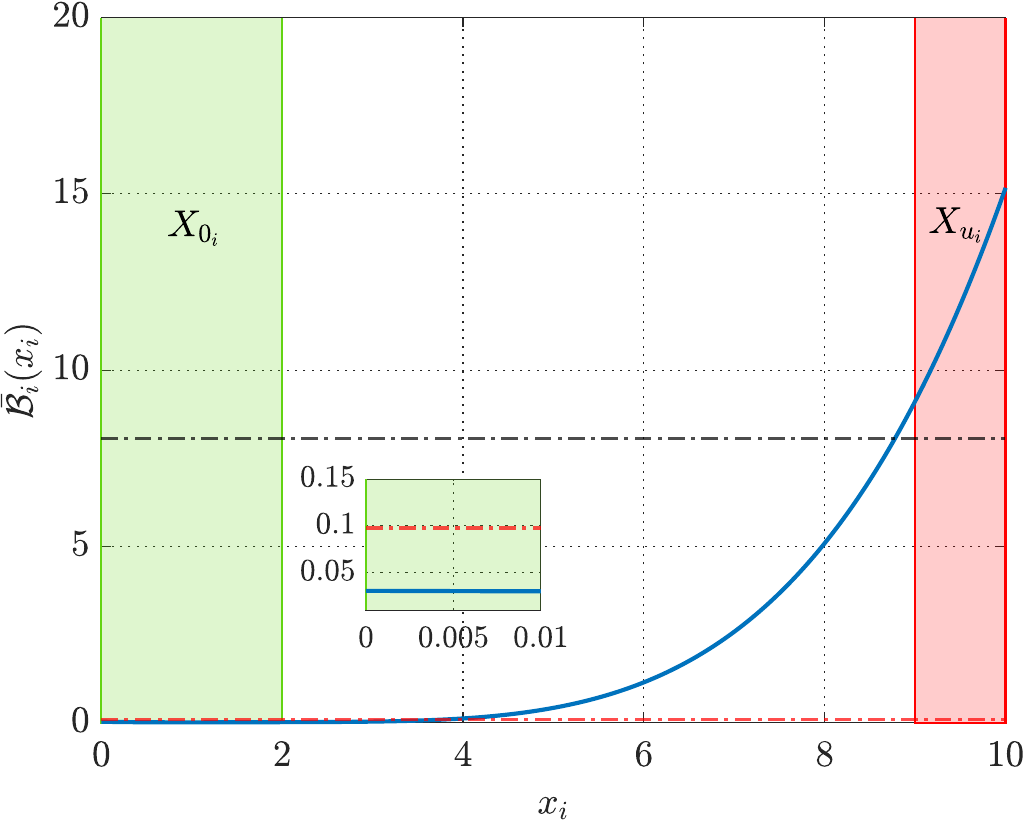}\\
	\includegraphics[width=0.3\linewidth]{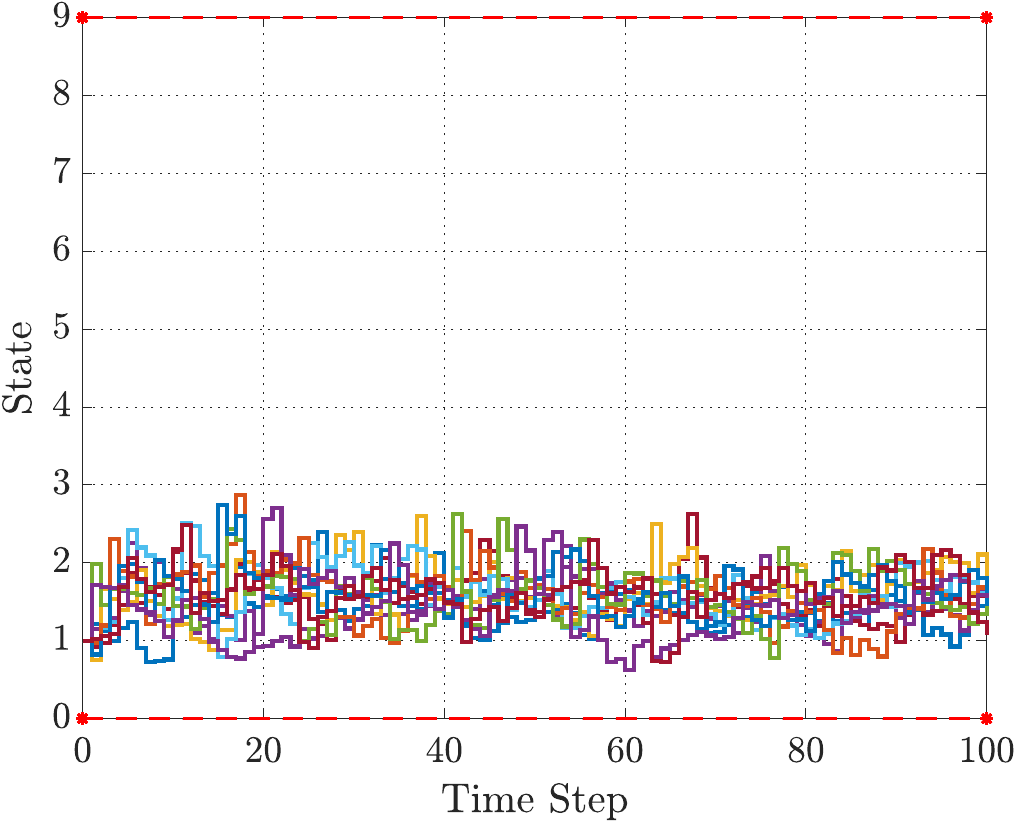}\hspace{0.5cm}
	\includegraphics[width=0.3\linewidth]{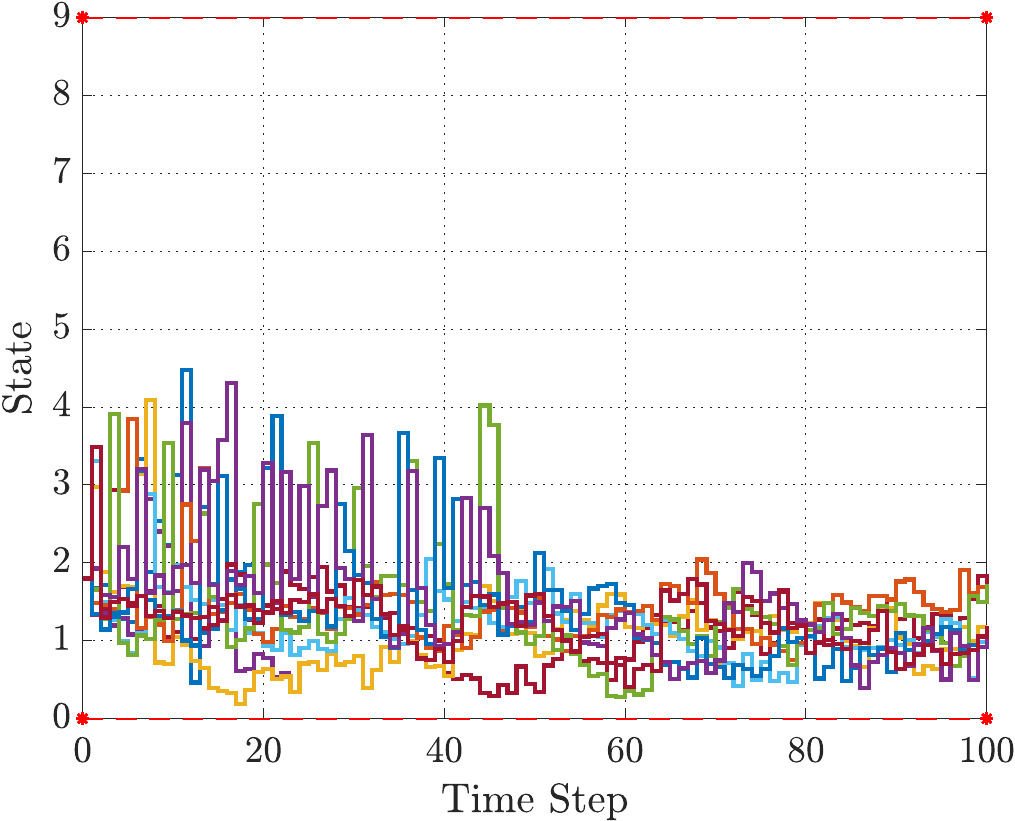}\hspace{0.5cm}
	\includegraphics[width=0.3\linewidth]{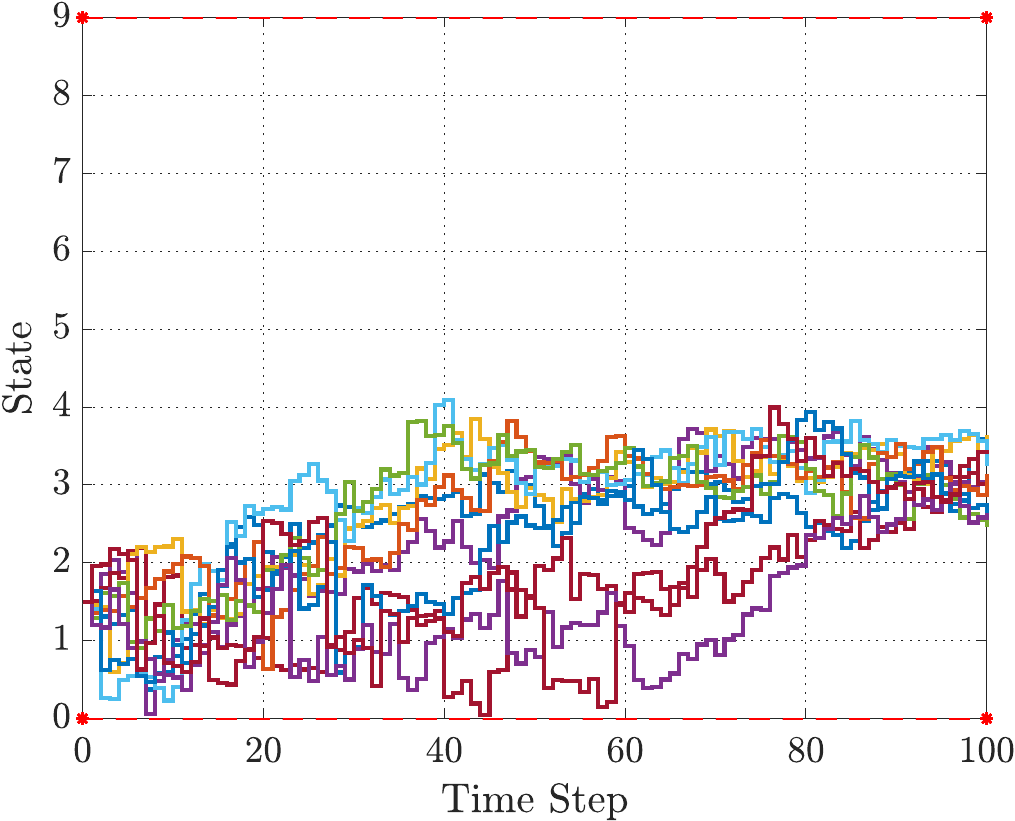}
	\caption{Control barrier certificate \sampleline{mycolor, thick} (top) and closed-loop state trajectories (bottom) of a representative subsystem of the interconnected SHS $\Phi$ with 3-in-3-out topology, under 10 noise realizations, for  $a_1 = -0.4,b_1 = 0.5,a_2 = 0.01,b_2 = 0.6$ (left), $a_1 = -0.3,b_1 = 0.2,a_2 = 1.01,b_2 = 1$ (middle), $a_1 = 0.01,b_1 = 0.7,a_2 = 0.075,b_2 = 0.8$ (right). Initial and unsafe regions are depicted by green \protect\greensquare\ and red \protect\redsquare\ boxes. Moreover, $\bar {\mathcal B}_i(x_i) = \bar {\mu}_i$ and $\bar {\mathcal B}_i(x_i) = \bar {\beta}_i$ are indicated by~\sampleline{dash dot, red, thick} and~\sampleline{dash dot, black , thick}, respectively. {Jump instants are assumed to be identical among all subsystems.}}
	\label{fig:w3}
\end{figure*}

For $a_1 = -0.3$, $b_1 = 0.2$, $a_2 = 1.01$, and $b_2 = 1$, we construct a CBC and its safety controller as $\bar{\mathcal B}_i(x_i) = 0.00095x^4_i - 0.0079x^3_i + 0.0233x^2_i - 0.0289x_i + 0.019$ and $\nu_i = -1.01x^3_i + 3$, respectively. The parameters satisfying~\eqref{subsys21}--\eqref{csbceq} in Definition~\ref{cbc} are attained as $\bar\mu_i = 0.0198$, $\bar\beta_i = 2.0239$, $\bar\kappa_i =  4.8932 \times 10^{-4}$, $\bar\eta_{1_i} = 0.031$, $\bar\eta_{2_i} = 0.003$, $\bar\varphi_{1_i} = 3.2398 \times 10^{-6}$, $\bar\varphi_{2_i} = 1.4533 \times 10^{-6}$, $\gamma_{1_i} = 0.951$, and $\gamma_{2_i} = 1.00001$. By Theorem~\ref{Thm:Main}, since $\gamma_{1_i} > 0$ and $\gamma_{2_i} \geq 1$, the results of Bound 2 are useful. Consequently, as condition~\eqref{Eq:12} is satisfied for all $\vartheta_i\in\{1,\ldots,7\}$, we obtain $\mathcal B_i(x_i,\vartheta_i) \coloneq e^{9.5 \times 10^{-3} \vartheta_i} \bar{\mathcal B}_i(x_i)$ as an A-CSBC for each $\mathcal A_i(\Phi_i)$, with $\mu_i = 0.0198$, $\beta_i = 2.0239$, $\gamma_i =\max\{e^{-\gamma_{1_i}\tau(1-\alpha_{1_i})},e^{-\gamma_{1_i}\tau \alpha_{1_i} \varepsilon_{1_i}}\gamma_{2_i}\}=\max\{ 0.9180,    0.9905 \} =0.9905$, $\varphi_i =\max\{e^{\gamma_{1_i}\tau \alpha_{1_i} \varepsilon_{2_i}}$ $\tfrac{(1-e^{-\gamma_{1_i}\!\tau})}{\gamma_{1_i}} \bar\varphi_{1_i},\bar\varphi_{2_i}\}=\max\{ 3.3032 \times 10^{-7}, 1.4533 \times 10^{-6}\} \!=\! 1.4533 \times 10^{-6}$, and $\eta_i \!=\! \max\{e^{\gamma_{1_i}\tau \alpha_{1_i}\varepsilon_{2_i}}\tfrac{(1-e^{-\gamma_{1_i}\tau})}{\gamma_{1_i}}$ $\bar{\varphi}_{1_i}\,\|w_i\|_\infty^2 \!+\!e^{\gamma_{1_i}\tau \alpha_{1_i}\varepsilon_{2_i}}\tfrac{(1-e^{-\gamma_{1_i}\tau})}{\gamma_{1_i}}\bar{\eta}_{1_i},\,\bar{\eta}_{2_i}\}=\max\{0.0032,\\ 0.003\}= 0.0032$, by choosing $\alpha_{1_i} = 0.1$ and $\alpha_{2_i} = 8$. From Theorem~\ref{Thm:2}, one can see $\xi_i \Gamma_i = -5.447 \times 10^{-7} < 0$ fulfills condition~\eqref{Eq:21}, with $\xi_i = 0.001$, $\forall i\in \{1,\dots,1000\}$. Moreover, $\beta =\sum_{i=1}^{1000}\xi_i \beta_i = 2.1632$ and $\mu = \sum_{i=1}^{1000}\xi_i \mu_i= 0.02$ satisfy condition~\eqref{Eq:28}. Thereby, the A-CBC $\mathcal B(x,\vartheta) = \sum_{i=1}^{1000} \xi_i \mathcal B_i(x_i,\vartheta_i)$ satisfies conditions~\eqref{Eq_2a1}--\eqref{Eq_3a} with $\eta = \sum_{i=1}^{1000}\xi_i\eta_i = 0.0032$ and $\gamma =1 + \Gamma= 0.9995$.

For $a_1 = 0.01$, $b_1 = 0.7$, $a_2 = 0.075$, and $b_2 = 0.8$, we synthesize a CBC and its corresponding safety controller as $\bar{\mathcal B}_i(x_i) = 0.0031x^4_i - 0.0209x^3_i + 0.0522x^2_i - 0.0524x_i + 0.0308$ and $\nu_i = -0.103x^3_i + 1.2$. The parameters satisfying~\eqref{subsys21}--\eqref{csbceq} in Definition~\ref{cbc} are $\bar\mu_i =   0.0978$, $\bar\beta_i = 8.073$, $\bar\kappa_i =     0.0032$, $\bar\eta_{1_i} = 0.01$, $\bar\eta_{2_i} = 0.01$, $\bar\varphi_{1_i} =  2.921 \times 10^{-5}$, $\bar\varphi_{2_i} = 9.4132\times 10^{-6}$, $\gamma_{1_i} = -0.21$, and $\gamma_{2_i} = 0.77$. As $\gamma_{1_i} \leq 0$ and $0 < \gamma_{2_i} < 1$, and given that condition~\eqref{Eq:12} is satisfied for all $\vartheta_i\in\{1,\ldots,7\}$, Bound 3 in Theorem~\ref{Thm:Main}, with $\alpha_{1_i} = 0.1$ and $\alpha_{2_i} = 8$, guarantees that $\mathcal B_i(x_i,\vartheta_i) \coloneq 0.77^{\frac{\vartheta_i}{8}} \bar{\mathcal B}_i(x_i)$ is an A-CSBC for each $\mathcal A_i(\Phi_i)$, with $\mu_i = 0.0978$, $\beta_i = 6.4227$, $\gamma_i =\max\{e^{-\gamma_{1_i}\tau}\gamma_{2_i}^{\frac{1}{\alpha_{2_i}}},\gamma_{2_i}^{\frac{\alpha_{2_i}-\varepsilon_{2_i}}{\alpha_{2_i}}}\}=\max\{0.9884,0.9679\} = 0.9884$, $\varphi_i = \max\{\gamma_{2_i}^{\frac{1}{\alpha_{2_i}}}e^{-\gamma_{1_i}\tau} \tau\bar\varphi_{1_i},\bar\varphi_{2_i}\}=\max\{ 2.8871 \times 10^{-6}, 9.4132 \times 10^{-6}\}= 9.4132 \times 10^{-6}$, and $\eta_i =\max\{\gamma_{2_i}^{\frac{1}{\alpha_{2_i}}}e^{-\gamma_{1_i}\tau} \tau\bar\varphi_{1_i}\Vert w_i\Vert_\infty^2 + \gamma_{2_i}^{\frac{1}{\alpha_{2_i}}}e^{-\gamma_{1_i}\tau}\tau\bar\eta_{1_i},\bar\eta_{2_i}\}=\max\{ 0.0017,  0.01\}= 0.01$. Furthermore, condition~\eqref{Eq:21} is satisfied by $\xi_i \Gamma_i =  -5.2279 \times 10^{-7} < 0$, with $\xi_i = 0.001$, $\forall i\in \{1,\dots,1000\}$, and $\beta =\sum_{i=1}^{1000}\xi_i \beta_i =7.8135$ and $\mu =\sum_{i=1}^{1000}\xi_i \mu_i = 0.0778$ fulfill condition~\eqref{Eq:28}. Hence, according to Theorem~\ref{Thm:2}, $\mathcal B(x,\vartheta) = \sum_{i=1}^{1000} \xi_i \mathcal B_i(x_i,\vartheta_i)$ is an A-CBC for $\mathcal A(\Phi)$ and satisfies conditions~\eqref{Eq_2a1}--\eqref{Eq_3a} with $\eta =\sum_{i=1}^{1000}\xi_i \eta_i = 0.01$, and $\gamma =1+\Gamma= 0.9995$.

According to Theorem~\ref{Kushner} and Proposition~\ref{Proposition}, the state trajectory of $\Phi$, initialized within $X_0 = [0,2]^{1000}$, {does not enter the unsafe set $X_u=[9,10]^{1000}$} over the time horizon $\mathcal T = 100$ with probabilities of at least $0.9678$ (Bound~1), $0.8435$ (Bound~2), and $0.8426$ (Bound~3)\footnote{Similarly, the reported safety probabilities were computed using full-precision values.}. Fig.~\ref{fig:w3} illustrates the synthesized CBC for a representative subsystem $\Phi_i$, along with the associated closed-loop state trajectories under Bounds~1--3 for 10 noise realizations, where only discrete time steps are shown, consistent with the guaranteed safety intervals. The CBC and corresponding controllers for each bound were computed in approximately $11.64$ seconds using $66.38~\mathsf{Mbit}$ of memory on a MacBook Pro with an M2 Max chip and 32 GB RAM.

\subsection{Comparison Study}
Comparing the results of our approach on a loosely coupled network, such as a ring topology, with those on a denser topology (\emph{e.g.,} a 3-in-3-out interconnection), reveals several emerging trends. First and foremost, satisfying compositional conditions~\eqref{Eq:21}--\eqref{Eq:28} becomes more challenging in all three cases as the value of $\xi_i \Gamma_i$ increases, \emph{i.e.}, it approaches zero. This arises from the fact that an increase in the dimension of the adversarial input $w_i$ pushes the matrix $(-\Theta + \Psi)$ closer to the positive semidefinite cone. Second, the safety probability associated with each bound \emph{decreases} in the 3-in-3-out interconnection topology compared to its counterpart in the ring topology (with the same certificate template, \emph{i.e.}, a CBC of degree four). Higher safety probabilities could potentially be achieved by employing a more expressive template with a higher degree; however, this would again lead to increased computational complexity. Last but not least, both computation time and memory usage are higher for the denser interconnection topology. This is mainly due to the fact that, according to Lemma~\ref{sos}, the number of coefficients in the SOS polynomials $l_{w_i}(x_i, \nu_i, w_i)$ and $\hat{l}_{w_i}(x_i, \nu_i, w_i)$ increases since higher connectivity directly leads to a higher-dimensional $w_i$, thereby amplifying the computational burden at the subsystem level.
\subsection{Computational Complexity Analysis}\label{subsec:complexity}
Here, we aim to {provide an intuitive discussion} about the computational complexity of our proposed approach compared with the existing method in the relevant literature~\citep{lavaei2022safety}. {In this respect, while the number of coefficients to be designed grows polynomially in the monolithic setting with respect to the total system size~\citep{lavaei2022safety}, the proposed compositional framework reduces it to the subsystem-level size. To illustrate this, we consider a network composed of scalar subsystems with a ring interconnection topology (similar to Section~\ref{subsec:ring} but with heterogeneous subsystems). For simplicity, in the sequel, we focus on the number of coefficients to be designed for the CBC, while emphasizing that analogous arguments can also be applied to the coefficients of the Lagrangian multipliers. Specifically, when solving a safety problem by searching for a polynomial CBC using \textsf{SOSTOOLS} in a monolithic fashion, for a fixed degree $d$ of the CBC (\emph{e.g.,} CBC of degree four, as in our case study) and an interconnected network of total size $s = n_i \times N$, the number of coefficients to be designed is $\binom{s + d}{s}$. Consequently, the number of coefficients to be designed grows polynomially with $s$ (and consequently $N$) for a fixed $d$. Accordingly, the associated computational complexity, characterized by the growth in the number of CBC coefficients, denoted by $\mathcal{C}$, roughly grows as $\mathcal{C}\propto\mathcal O(s^d)$.}

{As a result, the monolithic approach in~\citep{lavaei2022safety} requires solving the problem in our case study in Section~\ref{subsec:ring} (but with heterogeneous subsystems) for a monolithic system of dimension $1000$ (corresponding to $1000$ one-dimensional subsystems), leading to a computational cost on the order of $\mathcal{C} \propto \mathcal{O}(1000^4)$, which is computationally prohibitive. In contrast, our method decomposes the network into $1000$ one-dimensional subsystems, solves the problem individually for each, and subsequently lifts the safety guarantees over the entire network using compositional reasoning. Even though the per-subsystem complexity, denoted by $\bar{\mathcal{C}}$, still remains polynomial, {\emph{i.e.},} $\bar{\mathcal{C}}\propto\mathcal O(n_i^d)$, the overall complexity scales linearly in this example with the number of subsystems, yielding a total cost of $\mathcal C \propto \mathcal O(N)$, or roughly $\mathcal C \propto N\bar{\mathcal{C}}$, which is $\mathcal C \propto 1000\bar{\mathcal{C}}$. It is worth noting that, for fully interconnected networks, the computational complexity of our approach can become comparable to that of the monolithic method.}

\begin{figure}[t!]
	\centering
	\includegraphics[width=0.7\linewidth]{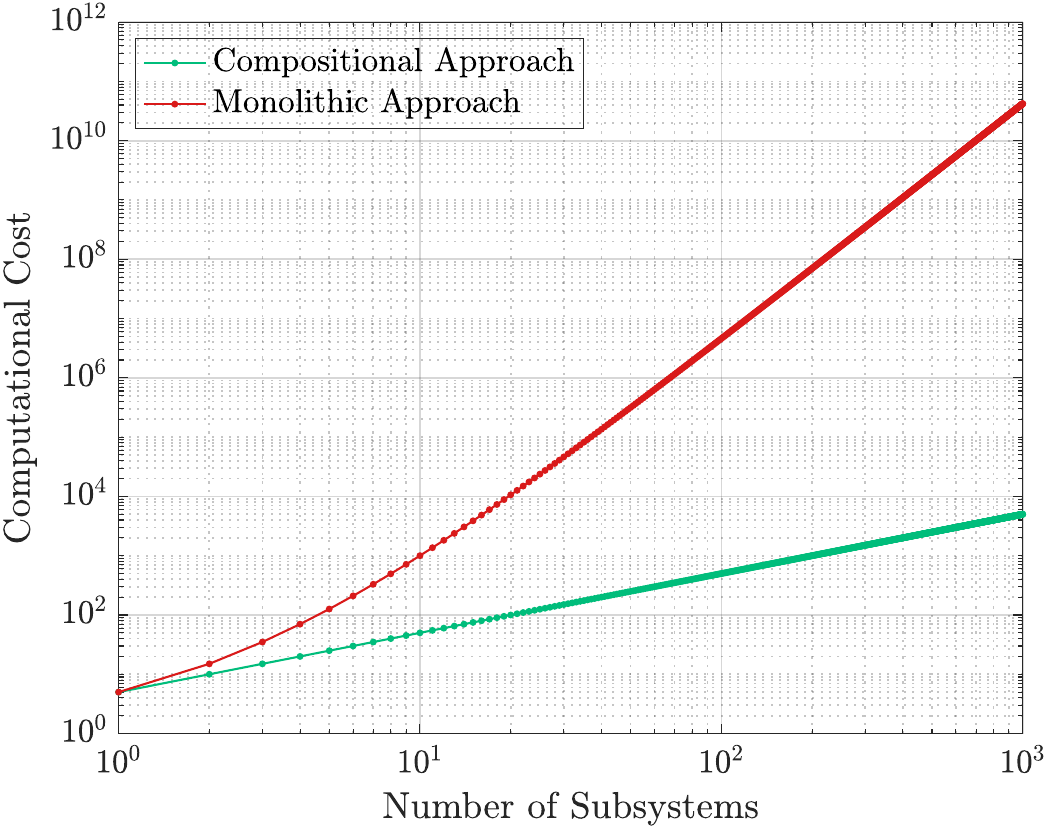}
	\caption{Comparison of computational complexity, based on the growth in the number of CBC coefficients, for the monolithic approach~\citep{lavaei2022safety} and the proposed compositional approach.}
	\label{fig:complexity}
\end{figure}
\begin{remark}
In principle, the underlying computations can be performed in parallel, whereby the overall computation time becomes independent of the number of subsystems and depends only on the state dimension, given sufficient computational resources. For instance, solving the safety problem for a network of $1000$ subsystems in parallel would require either $1000$ individual processing units ({e.g.}, personal computers) or a single high-performance computing platform ({e.g.}, a GPU-equipped server).
\end{remark}
\section{Conclusion}\label{Conclude}
In this work, we proposed a compositional framework for the safety controller synthesis of interconnected stochastic hybrid systems with both continuous evolution and instantaneous jumps. In our proposed setting, we first introduced an augmented framework {to reformulate each stochastic hybrid subsystem within a unified framework that captures both continuous and discrete evolutions.} We then introduced notions of augmented {control sub-barrier certificates}, constructed for each subsystem, by employing which we constructed augmented {control barrier certificates} for interconnected systems together with their safety controllers under some small-gain compositional conditions. By utilizing the constructed augmented control barrier certificate, we computed a guaranteed probabilistic bound on the safety of interconnected stochastic hybrid systems.

\section*{Acknowledgment}
The authors would like to thank Abdalla Swikir for fruitful discussions on the augmented system, and Amy Nejati for her insights on Lemma~\ref{Lemma:1}.

\bibliographystyle{agsm}
\bibliography{biblio}

\begin{authorbio}[Mahdieh]{Mahdieh Zaker} received her B.Sc. from K. N. Toosi University of Technology, Tehran, Iran, in 2019, and her M.Sc. from Amirkabir University of Technology (Tehran Polytechnic), Tehran, Iran, both in Electrical Engineering, control major. She is currently a PhD student in the School of Computing at Newcastle University, UK. She is the Best Repeatability Prize Finalist at the $28^{\text{th}}$ ACM International Conference on Hybrid Systems: Computation and Control (HSCC), 2025. Her research interests are (nonlinear) control and systems theory, data-driven techniques, large-scale systems, and formal methods.\vspace{-0.7cm}
\end{authorbio}

\begin{authorbio}[Omid]{Omid Akbarzadeh} is currently a PhD student in the School of Computing at Newcastle University, U.K. His academic journey commenced at Shiraz University, where he obtained a Bachelor of Science in Electrical and Electronic Engineering. Following this, he pursued a master's degree in Communications and Computer Network Engineering (CCNE) at the Polytechnic University of Turin, Italy (Politecnico di Torino). His research interests include safe cyber-physical systems, communication networks, data-driven approaches, and formal control.\vspace{-0.8cm}
\end{authorbio}

\begin{authorbio}[Behrad]{Behrad Samari} received his B.Sc. and M.Sc. degrees in electrical engineering, control major, from K. N. Toosi University of Technology, Tehran, Iran, and University of Tehran (UT), Tehran, Iran, in 2019 and 2022, respectively. He is currently pursuing his PhD in the School of Computing at Newcastle University, U.K. He is the Best Repeatability Prize Finalist at the 8$^{\text{th}}$ IFAC Conference on Analysis and Design of Hybrid Systems (ADHS), 2024. His research interests include (nonlinear) control and system theory, data-driven approaches, and formal methods.\vspace{-0.6cm}
\end{authorbio}

\begin{authorbio}[Abolfazl]{Abolfazl Lavaei} is an Assistant Professor in the School of Computing at Newcastle University, United Kingdom. Between January 2021 and July 2022, he was a Postdoctoral Associate in the Institute for Dynamic Systems and Control at ETH Zurich, Switzerland. He was also a Postdoctoral Researcher in the Department of Computer Science at LMU Munich, Germany, between November 2019 and January 2021. He received the Ph.D. degree in Electrical Engineering from the Technical University of Munich (TUM), Germany, in 2019. He obtained the M.Sc. degree in Aerospace Engineering with specialization in Flight Dynamics and Control from the University of Tehran (UT), Iran, in 2014. He is the recipient of several international awards in the acknowledgment of his work including  Best Repeatability Prize (Finalist) at the ACM HSCC 2025, IFAC ADHS 2024, and IFAC ADHS 2021, HSCC Best Demo/Poster Awards 2022 and 2020, IFAC Young Author Award Finalist 2019, and Best Graduate Student Award 2014 at University of Tehran with the full GPA (20/20). His research interests revolve around the intersection of Control Theory, Formal Methods in Computer Science, and Statistical Learning Theory.
\end{authorbio}

\end{document}